\documentclass[aps,prb,superscriptaddress,twocolumn,showpacs,preprintnumbers,amsmath,amssymb]{revtex4}
\usepackage{color}
\usepackage{array}
\usepackage{amsmath}
\usepackage{amssymb}
\usepackage{epsfig}
\usepackage{graphicx}
\usepackage{bbm}
\usepackage{hyperref}

\newcommand{\Ref}[1]{Ref.~\onlinecite{#1}}
\newcommand{\bs}[1]{{\boldsymbol#1}}
\newcommand{\bss}{{\boldsymbol{\sigma}}}
\newcommand{\bst}{{\boldsymbol{T}}}
\newcommand{\bse}{{\boldsymbol{e}}}
\newcommand{\ie}{{\emph{i.e.~}}}
\makeatletter
\newcommand{\rmnum}[1]{\romannumeral #1}
\newcommand{\Rmnum}[1]{\expandafter\@slowromancap\romannumeral #1@}
\makeatother
\newcommand{\imth}{\hspace{1pt}\mathrm{i}\hspace{1pt}}
\newcommand{\alert}[1]{{\color{red}{#1}}}
\newcommand{\eg}{{\emph{e.g.~}}}

\newcommand{\cs}{{C_6}}
\newcommand{\tk}{\textbf{k}}
\newcommand{\kag}{\text{kagome}}

\begin{document}
\title{$Z_2$ spin liquids in $S=1/2$ Heisenberg model on $\kag$ lattice: projective symmetry group study of Schwinger-fermion mean-field states}

\author{Yuan-Ming Lu}\author{Ying Ran}
\affiliation{Department of Physics, Boston College, Chestnut Hill,
Massachusetts, 02467, USA}
\author{Patrick A. Lee}
\affiliation{Department of Physics, Massachusetts Institute of
Technology, Cambridge, Massachusetts, 02139, USA}
\date{\today}

\begin{abstract}
Due to strong geometric frustration and quantum fluctuation, $S=1/2$
quantum Heisenberg antiferromagnets on the $\kag$ lattice has long
been considered as an ideal platform to realize spin liquid (SL), a
novel phase exhibiting fractionalized
excitations without any symmetry breaking. A recent numerical study[\onlinecite{Yan2010}] of
Heisenberg $S=1/2~\kag$ lattice model (HKLM) shows that in contrast
to earlier results, the ground state is a singlet-gapped SL with
signatures of $Z_2$ topological order. Motivated by this numerical
discovery, we use projective symmetry group to classify all 20
possible Schwinger-fermion mean-field states of $Z_2$ SLs on $\kag$
lattice. Among them we found only one gapped $Z_2$ SL (which we call
$Z_2[0,\pi]\beta$ state) in the neighborhood of $U(1)$-Dirac SL
state. Since its parent state, \ie $U(1)$-Dirac SL is found[\onlinecite{Ran2007}] to be the lowest
among many other candidate $U(1)$ SLs including the uniform
resonating-valence-bond states, we propose this
$Z_2[0,\pi]\beta$ state to be the numerically discovered SL ground
state of HKLM.
\end{abstract}


\maketitle

\section{Introduction}

At zero temperature all degrees of freedom tend to freeze and usually a
variety of different orders, such as
superconductivity and magnetism, will develop in different materials. However, in a quantum system with a large zero-point energy, one may expect a
liquid-like ground state to exist even at $T=0$. In a system consisting of
localized quantum magnets, we call such a quantum-fluctuation-driven
disordered ground state a quantum spin liquid (SL)\cite{Lee2008}. It
is an exotic phase with novel ``fractionalized" excitations carrying
only a fraction of the electron quantum number, \eg spinons which
carry spin but no charge. The internal structures of these SLs are
so rich that they are beyond the description of Landau's symmetry
breaking theory\cite{Landau1937} of conventional ordered phases.
Instead they are characterized by long-range quantum
entanglement\cite{Levin2006,Kitaev2006a} encoded in the ground
state, which is coined ``topological order"\cite{Wen1990,Wen2002} in
contrast to the conventional symmetry-breaking order.

Geometric frustration in a system of quantum magnets would lead to a huge
degeneracy of classical ground state configurations. The
quantum tunneling among these classical ground states provides a mechanism to realize quantum SLs. The quest for quantum SLs in
frustrated magnets (for a recent review see \Ref{Balents2010}) has
been pursued for decades. Among them the Heisenberg $S=1/2$ kagome
lattice model (HKLM)
\begin{eqnarray}\label{hklm}
H_{HKLM}=J\sum_{<i,j>}\bs{S}_i\cdot\bs{S}_j
\end{eqnarray}
has long been thought as a promising candidate. Here $<i,j>$
denotes $i,j$ being a nearest neighbor pair. Experimental evidence of
SL\cite{Mendels2007,Helton2007,Imai2008,Helton2010} has been
observed in ZnCu$_3$(OH)$_6$Cl$_2$ (called herbertsmithite), a
spin-half antiferromagnet on the $\kag$ lattice. Theoretically, in
lack of an exact solution of the two-dimensional (2D) quantum
Hamiltonian (\ref{hklm}) in the thermodynamic limit, in previous
studies either a honeycomb valence bond
crystal\cite{Marston1991,Nikolic2003,Singh2007,Singh2008,Evenbly2010}
(HVBC) with an enlarged $6\times6$-site unit cell, or a gapless
SL\cite{Jiang2008} were proposed as the ground state of HKLM.
However, recently an extensive density-matrix-renormalization-group
(DMRG) study\cite{Yan2010} on HKLM reveals the ground state of HKLM
as a gapped SL, which substantially lowers the energy compared to HVBC. Besides, they also observe numerical signatures of
$Z_2$ topological order in the SL state.

\begin{figure}
 \includegraphics[width=0.22\textwidth]{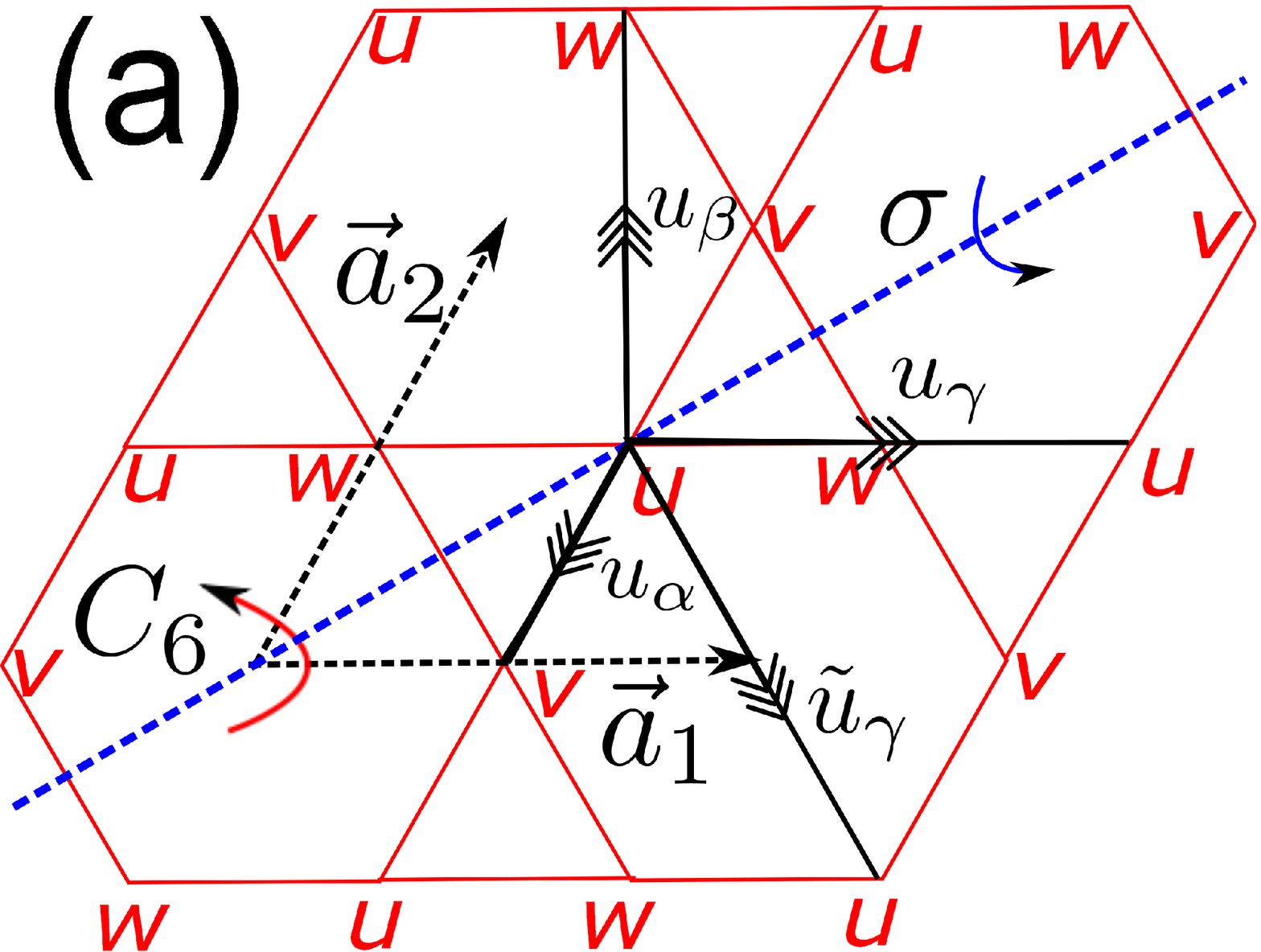}\;\includegraphics[width=0.22\textwidth]{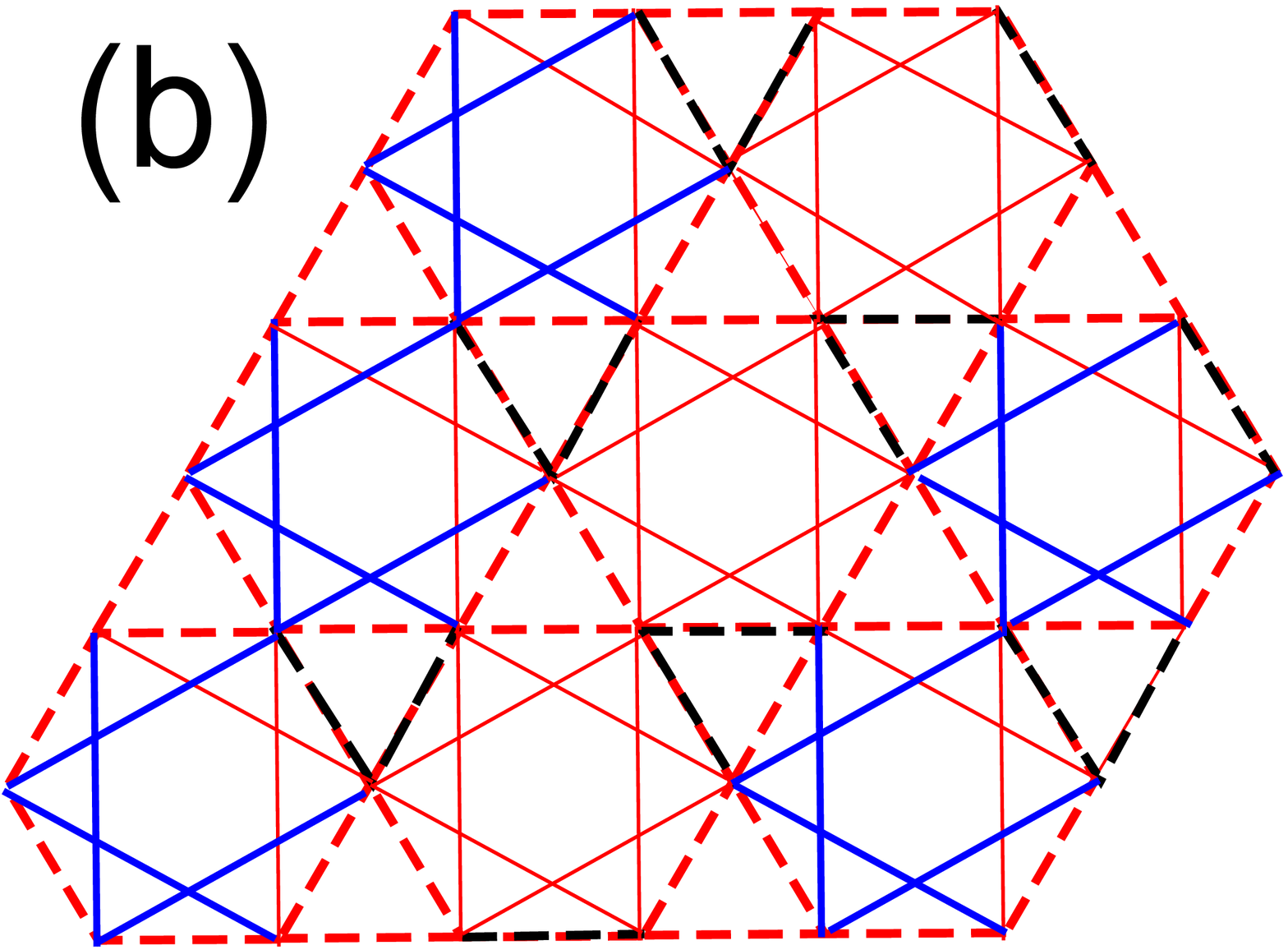}
\caption{(color online) (a) $\kag$ lattice and the elements of its
symmetry group. $\vec a_{1,2}$ are the translation unit vectors,
$C_6$ denotes $\pi/3$ rotation around honeycomb center and $\bss$
represents mirror reflection along the dashed blue line. Here
$u_\alpha$ and $u_\beta$ denote 1st and 2nd nearest neighbor (n.n.)
mean-field bonds while $u_\gamma$ and $\tilde u_\gamma$ represent
two kinds of independent 3rd n.n. mean-field bonds. (b) Mean-field
ansatz of $Z_2[0,\pi]\beta$ state up to 2nd nearest neighbor. Colors
in general denote the sign structure of mean-field bonds. Dashed
lines denote 1st n.n. real hopping terms
$\chi_1\sum_{<i,j>\alpha}(\nu_{ij}f^\dagger_{i\alpha}f_{j\alpha}+~h.c.)$:
red ones have $\nu_{ij}=1$ and black ones have $\nu_{ij}=-1$. Solid
lines stand for 2nd n.n. hopping
$\chi_2\sum_{<<ij>>\alpha}\nu_{ij}(f^\dagger_{i\alpha}f_{j\alpha}+~h.c.)$
and singlet pairing
$\sum_{<<ij>>\alpha\beta}\epsilon_{\alpha\beta}\nu_{ij}(\Delta_2f^\dagger_{i\alpha}f^\dagger_{j\beta}+~h.c.)$:
again red ones have $\nu_{ij}=1$ and blue ones have $\nu_{ij}=-1$.
Here $\chi_{1,2}$ and $\Delta_2$ are real parameters after choosing
a proper gauge. } \label{fig:kagome}
\end{figure}

Motivated by this important numerical discovery, we try to find out
the nature of this gapped $Z_2$ SL. Different $Z_2$ SLs on the
$\kag$ lattice have been previously studied using Schwinger-boson
representation\cite{Sachdev1992,Wang2006}. Here we propose the candidate
states of symmetric $Z_2$ SLs on $\kag$ lattice by Schwinger-fermion
mean field
approach\cite{Baskaran1987,Affleck1988a,Baskaran1988,Kotliar1988,Mudry1994,Wen1996,Lee2006}.
Following is the summary of our results. First we use projective
symmetry group\cite{Wen2002} (PSG) to classify all 20 possible
Schwinger-fermion mean-field ansatz of $Z_2$ SLs which preserve all
the symmetry of HKLM, as shown in TABLE \ref{tab:MF_ansatz}. We
analyze these 20 states and rule out some obviously unfavorable
states: \eg gapless states, and those states whose 1st nearest
neighbor (n.n.) mean-field amplitudes must vanish due to symmetry. Then we focus on those $Z_2$ SLs in the neighborhood of
the $U(1)$-Dirac SL\cite{Ran2007}. In \Ref{Ran2007}~it is shown that
$U(1)$-Dirac SL has a significantly lower energy compared with other
candidate $U(1)$ SL states, such as the uniform
resonating-valence-bond (RVB) state(or the $U(1)$ SL-$[0,0]$ state
in notation of \Ref{Ran2007}). We find out that there is only one
gapped $Z_2$ SL, which we label as $Z_2[0,\pi]\beta$, in the
neighborhood of (or continuously connected to) $U(1)$-Dirac SL. Therefore we propose this
$Z_2[0,\pi]\beta$ state as a promising candidate state for the
ground state of HKLM. The mean-field ansatz of $Z_2[0,\pi]\beta$
state is shown in FIG. \ref{fig:kagome}(b). Our work also provides
guideline for choosing variational states in future numeric studies
of SL ground state on $\kag$ lattice.

\section{Schwinger-fermion construction of spin liquids and projective symmetry group (PSG)}

\subsection{Schwinger-fermion construction of symmetric spin liquids}

In the Schwinger-fermion
construction\cite{Baskaran1987,Affleck1988a,Baskaran1988,Kotliar1988,Mudry1994,Wen1996},
we represent a spin-1/2 operator at site $i$ by fermionic spinons
$\{f_{i\alpha},~\alpha=\uparrow,\downarrow\}$:
\begin{align}
\vec
S_i=\frac{1}{2}f_{i\alpha}^{\dagger}\vec\sigma_{\alpha\beta}f_{i\beta}.\label{eq:schwinger-fermion}
\end{align}
Heisenberg hamiltonian $H=\sum_{<ij>} J_{ij} \vec S_i\cdot \vec S_j$
is represented as
$H=\sum_{<ij>}-\frac{1}{2}J_{ij}\big(f_{i\alpha}^{\dagger}f_{j\alpha}f_{j\beta}^{\dagger}f_{j\beta}+\frac{1}{2}f_{i\alpha}^{\dagger}f_{i\alpha}f_{j\beta}^{\dagger}f_{j\beta}\big)$.
This construction
enlarges the Hilbert space of the original spin system. To obtain
the physical spin state from a mean-field state of $f$-spinons, we
need to enforce the following one-$f$-spinon-per-site constraint:
\begin{align}
 f_{i\alpha}^{\dagger}f_{i\alpha}&=1,&f_{i\alpha}f_{i\beta}\epsilon_{\alpha\beta}=0.\label{eq:constraint}
\end{align}
Mean-field parameters of symmetric SLs are
$\Delta_{ij}\epsilon_{\alpha\beta}=-2\langle
f_{i\alpha}f_{j\beta}\rangle$,
$\chi_{ij}\delta_{\alpha\beta}=2\langle
f_{i\alpha}^{\dagger}f_{j\beta}\rangle$, where
$\epsilon_{\alpha\beta}$ is the completely antisymmetric tensor. Both terms
are invariant under global $SU(2)$ spin rotations. After a Hubbard-Stratonovich transformation, the lagrangian of the spin
system can be written as
\begin{align}
L=&\sum_i\psi_{i}^{\dagger}\partial_{\tau}\psi_{i}+\sum_{<ij>}\frac{3}{8}J_{ij}\big[\frac{1}{2}\mbox{Tr}(U^{\dagger}_{ij}U_{ij})\notag\\
&-(\psi^{\dagger}_i U_{ij}\psi_{j}+h.c.)\big]+\sum_i a_0^l(i)
\psi_i^{\dagger}\tau^l\psi_i\label{eq:action}
\end{align}
where two-component fermion notation
$\psi_i\equiv(f_{i\uparrow},f_{i\downarrow}^{\dagger})$ is introduced for
reasons that will be explained shortly. We use $\tau^0$ to denote
the $2\times2$ identity matrix and $\tau^{1,2,3}$ are the three
Pauli matrices. $U_{ij}$ is a matrix of mean-field amplitudes:
\begin{align}
 U_{ij}=\begin{pmatrix}\chi_{ij}^{\dagger}&\Delta_{ij}\\ \Delta_{ij}^{\dagger}&-\chi_{ij}\end{pmatrix}.
\end{align}
$a_0^{l}(i)$ are the local lagrangian multipliers that enforce the
constraints Eq.(\ref{eq:constraint}).

In terms of $\psi$, Schwinger-fermion representation has an explicit
$SU(2)$ gauge redundancy: a transformation $\psi_i\rightarrow
W_i\psi_i$, $U_{ij}\rightarrow W_i U_{ij}W_j^{\dagger}$, $W_i\in
SU(2)$ leaves the action invariant. This redundancy is originated
from representation Eq.(\ref{eq:schwinger-fermion}): this local
$SU(2)$ transformation leaves the spin operators invariant and
does not change physical Hilbert space. One can try to solve
Eq.(\ref{eq:action}) by mean-field (or saddle-point) approximation.
At mean-field level, $U_{ij}$ and $a_0^l$ are treated as complex
numbers, and $a_0^l$ must be chosen such that constraints
(\ref{eq:constraint}) are satisfied at the mean field level:
$\langle\psi_i^{\dagger}\tau^l \psi_i\rangle=0$. The mean-field
ansatz can be written as:
\begin{align}
 H_{MF}=-\sum_{<ij>}\psi^{\dagger}_i\langle i|j\rangle\psi_{j}+\sum_{i}\psi_i^{\dagger}a_0^l\tau^l\psi_i.\label{eq:mf}
\end{align}
where we defined $\langle i|j\rangle\equiv\frac{3}{8}J_{ij}U_{ij}$.
Under a local $SU(2)$ gauge transformation $\langle
i|j\rangle\rightarrow W_i\langle i|j\rangle W_j^{\dagger}$, but the
physical spin state described by the mean-field ansatz $\{\langle
i|j\rangle\}$ remains the same. By construction the mean-field
ansatz does not break spin rotation symmetry, and the mean field
solutions describe SL states if lattice symmetry is preserved.
Different $\{\langle i|j\rangle\}$ ansatz may be in different SL
phases. The mathematical language to classify different SL phases is
projective symmetry group (PSG)\cite{Wen2002}.

\subsection{Projective symmetry group (PSG) classification of topological orders in spin liquids}

PSG characterizes the topological order in Schwinger-fermion
representation: SLs described by different PSGs are different
phases. It is defined as the collection of all combinations of
symmetry group and $SU(2)$ gauge transformations that leave
mean-field ansatz $\{\langle i|j\rangle\}$ invariant (as $a_0^l$ are
determined self-consistently by $\{\langle i|j\rangle\}$, these
transformations also leave $a_0^l$ invariant). The invariance of a
mean-field ansatz $\{\langle i|j\rangle\}$ under an element of PSG
$G_U U$ can be written as
\begin{align}
G_U U(\{ \langle i|j\rangle\})&=\{\langle i|j\rangle\},\label{psg_definition}\\
U(\{\langle i|j\rangle\})&\equiv \{\tilde{\langle i|j\rangle}=\langle U^{-1}(i)|U^{-1}(j)\rangle\},\notag\\
G_U(\{\langle i|j\rangle\})&\equiv \{\tilde{\langle i|j\rangle}=G_U(i)\langle i|j\rangle G_U(j)^{\dagger}\}, \notag\\
&G_U(i)\in SU(2)\notag.
\end{align}
Here $U\in SG$ is an element of symmetry group (SG) of the
corresponding SL. In our case of symmetric SLs on the $\kag$
lattice, we use $(x,y,s)$ to label a site with sublattice index
$s=u,v,w$ and $x,y\in\mathbb{Z}$. Bravais unit vector are chosen as
$\vec a_1=a\hat x$ and $\vec a_2=\frac{a}2(\hat x+\sqrt3\hat y)$ as
shown in FIG. \ref{fig:kagome}(a). The symmetry group is generated
by time reversal operation $\bst$, lattice translations $T_{1,2}$
along $\vec a_{1,2}$ directions, $\pi/3$ rotation $\cs$ around
honeycomb plaquette center and the mirror reflection $\bss$ (for
details see Appendix \ref{app:symmetry group}). For example, if $U=T_1$ is the
translation along $\vec a_1$-direction in Fig.\ref{fig:kagome}(a),
$T_1(\{x,y,s\})=\{x+1,y,s\}$. $G_U$ is the gauge transformation
associated with $U$ such that $G_U U$ leave $\{\langle i|j\rangle\}$
invariant. Notice this condition (\ref{psg_definition}) allows us to
generate all symmetry-related mean-field bonds from one by the
following relation:
\begin{eqnarray}\label{psg:def}
\langle i|j\rangle=G_U(i)\langle U^{-1}(i)|U^{-1}(j)\rangle
G^\dagger_U(j)
\end{eqnarray}

There is an important subgroup of PSG, the invariant gauge group
(IGG), which is composed of all the pure gauge transformations in
PSG: $IGG\equiv\{\{W_i\}|W_i\langle i|j\rangle W_j^{\dagger}=\langle
i|j\rangle, W_i\in SU(2)\}$. In other words, $W_i=G_\bse(i)$ is the
pure gauge transformation associated with identity element $\bse\in
SG$ of the symmetry group. One can always choose a gauge in which
the elements in IGG is site-independent. In this gauge, IGG can be
the global $Z_2$ transformations: $\{G_\bse(i)\equiv
G_\bse=\pm\tau^0\}$, the global $U(1)$ transformations:
$\{G_\bse(i)\equiv e^{\imth\theta\tau^3},\theta\in[0,2\pi]\}$, or
the global $SU(2)$ transformations: $\{G_\bse(i)\equiv
e^{\imth\theta\hat n\cdot \vec\tau},\theta\in(0,2\pi],\hat n\in
S^2\}$, and we term them as $Z_2$, $U(1)$ and $SU(2)$ state
respectively.

The importance of IGG is that it controls the low energy gauge
fluctuations of the corresponding SL states. Beyond mean-field
level, fluctuations of $\langle i|j\rangle$ and $a_0^l$ need to be considered
and the mean-field state may or may not be stable. The low energy
effective theory is described by fermionic spinon band structure
coupled with a dynamical gauge field of IGG. For example, $Z_2$
state with gapped spinon dispersion can be a stable phase because
the low energy $Z_2$ dynamical gauge field can be in the deconfined
phase\cite{Wegner1971,Kogut1979}.


Notice that the condition $\{G_\bse(i)\equiv G_\bse=\pm\tau^0\}$ for
a $Z_2$ SL leads to a series of consistent conditions for the gauge
transformations $\{G_U(i)|U\in SG\}$, as shown in Appendix
\ref{app:symmetry group}. Gauge inequivalent solutions of these
conditions (\ref{algebra:psg:T})-(\ref{algebra:psg:C6}) lead to
different $Z_2$ SLs. Soon we will show that there are 20 $Z_2$ SLs
on the $\kag$ lattice that can be realized by a Schwinger-fermion mean-field ansatz
$\{\langle i|j\rangle\}$.

\section{$Z_2$ spin liquids on the kagome lattice and $Z_2[0,\pi]\beta$ state}

Following previous discussions, we use PSG to classify all possible
20 $Z_2$ SL states on $\kag$ lattice in this section. As will be
shown later, among them there is one gapped $Z_2$ SL labeled as
$Z_2[0,\pi]\beta$ state in the neighborhood of $U(1)$-Dirac SL. This $Z_2[0,\pi]\beta$ SL state is the most promising candidate for
the SL ground state of
HKLM.\\

\subsection{PSG classification of $Z_2$ spin liquids on $\kag$ lattice}

Applying the condition $G_\bse(i)\equiv G_\bse=\pm\tau^0$ to $\kag$
lattice with symmetry group described in Appendix \ref{app:symmetry
group}, we obtain a series of consistent conditions for the gauge
transformation $G_U(i)$, \ie conditions
(\ref{algebra:psg:T})-(\ref{algebra:psg:C6}). Solving these
conditions we classify all the 
20 different Schwinger-fermion mean-field states of $Z_2$ SLs on
$\kag$ lattice, as summarized in TABLE \ref{tab:MF_ansatz}. These 20
mean-field states correspond to different $Z_2$ SL phases, which
cannot be continuously tuned into each other without a phase
transition.

\begin{table}[tb]
\begin{tabular}{|c||c|c|c|c|c|c|c|c|c|c|}
\hline $\#$ & $\eta_{12}$&$\Lambda_s$&$u_\alpha$&$u_\beta$&$u_\gamma$&$\tilde u_\gamma$&Label&Gapped?\\
\hline 1&$+1$&$\tau^2,\tau^3$&$\tau^{2},\tau^3$&\alert{$\tau^2,\tau^3$}&$\tau^2,\tau^3$&$\tau^2,\tau^3$&$Z_2[0,0]A$&Yes\\
\hline {\bf2}&$-1$&$\tau^2,\tau^3$&$\tau^{2},\tau^3$&\alert{$\tau^2,\tau^3$}&$\tau^2,\tau^3$&$0$&$Z_2[0,\pi]\beta$&{\bf Yes}\\
\hline 3&$+1$&$0$&$\tau^{2},\tau^3$&$0$&$0$&$0$&$Z_2[\pi,\pi]A$&No\\
\hline 4&$-1$&$0$&$\tau^{2},\tau^3$&$0$&$0$&\alert{$\tau^2,\tau^3$}&$Z_2[\pi,0]A$&No\\
\hline 5&$+1$&$\tau^3$&\alert{$\tau^2,\tau^3$}&$\tau^3$&$\tau^3$&$\tau^3$&$Z_2[0,0]B$&Yes\\
\hline 6&$-1$&$\tau^3$&\alert{$\tau^2,\tau^3$}&$\tau^3$&$\tau^3$&$\tau^2$&$Z_2[0,\pi]\alpha$&No\\
\hline 7&$+1$&$0$&$0$&$\tau^2,\tau^3$&$0$&$0$&-&-\\
\hline 8&$-1$&$0$&$0$&$\tau^2,\tau^3$&$0$&$0$&-&-\\
\hline 9&$+1$&$0$&$0$&$0$&$\tau^2,\tau^3$&$0$&-&-\\
\hline 10&$-1$&$0$&$0$&$0$&$\tau^2,\tau^3$&$0$&-&-\\
\hline 11&$+1$&$0$&$0$&$\tau^2$&$\tau^2$&$0$&-&-\\
\hline 12&$-1$&$0$&$0$&$\tau^2$&$\tau^2$&$0$&-&-\\
\hline 13&$+1$&$\tau^3$&$\tau^3$&\alert{$\tau^2,\tau^3$}&$\tau^3$&$\tau^3$&$Z_2[0,0]D$&Yes\\
\hline 14&$-1$&$\tau^3$&$\tau^3$&\alert{$\tau^2,\tau^3$}&$\tau^3$&$0$&$Z_2[0,\pi]\gamma$&No\\
\hline 15&$+1$&$\tau^3$&$\tau^3$&$\tau^3$&\alert{$\tau^2,\tau^3$}&$\tau^3$&$Z_2[0,0]C$&Yes\\
\hline 16&$-1$&$\tau^3$&$\tau^3$&$\tau^3$&\alert{$\tau^2,\tau^3$}&$0$&$Z_2[0,\pi]\delta$&No\\
\hline 17&$+1$&$0$&$\tau^2$&\alert{$\tau^3$}&$0$&$0$&$Z_2[\pi,\pi]B$&No\\
\hline 18&$-1$&$0$&$\tau^2$&\alert{$\tau^3$}&$0$&$\tau^3$&$Z_2[\pi,0]B$&No\\
\hline 19&$+1$&$0$&$\tau^2$&$0$&$\tau^2$&$0$&$Z_2[\pi,\pi]C$&No\\
\hline 20&$-1$&$0$&$\tau^2$&$0$&$\tau^2$&\alert{$\tau^3$}&$Z_2[\pi,0]C$&No\\
\hline
\end{tabular}
\caption{\label{tab:MF_ansatz}Mean-field ansatz of 20 possible $Z_2$
SLs on a $\kag$ lattice. In our notation of mean-field amplitudes
$\langle x,y,s|0,0,u\rangle\equiv[x,y,s]$, this table summarizes all
symmetry-allowed mean-field bonds up to 3rd n.n., \ie
1st n.n. bond $u_\alpha=[0,0,v]$, 2nd n.n. bond $u_\beta=[0,1,w]$,
3rd n.n. bonds $u_\gamma=[1,0,u]$ and $\tilde u_\gamma=[1,-1,u]$ as
shown in FIG. \ref{fig:kagome}(a). $\Lambda_s$ denote the on-site
chemical potential terms which enforce the constraint
(\ref{constraint:mf}). $\tau^0$ is $2\times2$ identity matrix while $\tau^{1,2,3}$ are three Pauli matrices. $\tau^{0,3}$ denote hopping while
$\tau^{1,2}$ denote pairing terms. $0$ means the corresponding
mean-field amplitudes must vanish due to symmetry. Red color denotes
the shortest mean-field bonds necessary to realize a $Z_2$ SL. In
other words, the mean-field amplitudes with red color break the
$U(1)$ gauge redundancy down to $Z_2$ through Higgs mechanism. So in
$\#3,\#19$ and $\#7-\#12$ states a $Z_2$ SL cannot be realized with
up to 3rd n.n. mean-field amplitudes. Note that $\#15$ state needs
only 3rd n.n. bond $u_\gamma$ to realize a $Z_2$ SL ($\tilde
u_\gamma$ not necessary) , while state $\#20$ needs only $\tilde
u_\gamma$ to realize a $Z_2$ SL ($u_\gamma$ not necessary) . Notice
that when $\eta_{12}=-1$ the mean-field ansatz (instead of the SL
itself) will break translational symmetry and double the unit cell.
There are six $Z_2$ SLs, \ie $\#7-\#12$ that don't allow any 1st
n.n. mean-field bonds. Among the other 14 $Z_2$ SLs with
nonvanishing 1st n.n. mean-field bonds, only five $Z_2$ SL states,
\ie $\#1,\#2,\#5,\#13,\#15$ have gapped spinon spectra. $\#2$ or
$Z_2[0,\pi]\beta$ state in neighborhood of $U(1)$-Dirac SL is the most promising candidate of $Z_2$ SL
for the HKLM ground state.}
\end{table}

As discussed in Appendix \ref{app:MF_cond}, from PSG elements
$G_U(i)$ one can obtain all other symmetry-related mean-field bonds
from one using symmetry condition (\ref{psg:def}). Therefore we use
$u_\alpha\equiv\langle0,0,v|0,0,u\rangle$ to represent 1st nearest
neighbor (n.n.) mean-field bonds.
$u_\beta\equiv\langle0,1,w|0,0,u\rangle$ is the representative of 2nd
n.n. mean-field bonds. There are two kinds of symmetry-unrelated 3rd
n.n. mean-field bonds, represented by
$u_\gamma=\langle1,0,u|0,0,u\rangle$ and $\tilde
u_\gamma=\langle1,-1,u|0,0,u\rangle$. The symmetry conditions for
these mean-field bonds are summarized in
(\ref{condition:1st_nn})-(\ref{condition:3rd_nn2}). Besides, the
on-site chemical potential terms $\Lambda(i)$ (which guarantee the
physical constraint (\ref{eq:constraint}) on the mean-field level)
also satisfy symmetry conditions (\ref{condition:on-site}). We can
show that $\Lambda(x,y,s)\equiv\Lambda_s$ for these 20 $Z_2$ SL
states. The symmetry-allowed mean-field amplitudes/bonds are also
summarized in TABLE \ref{tab:MF_ansatz}.

From TABLE \ref{tab:MF_ansatz} we can see there are 6 states, \ie
$\#7-\#12$ that don't allow nonzero 1st n.n. mean-field amplitudes due to
symmetry. Moreover, they cannot realize $Z_2$ SLs with up to 3rd
n.n. mean-field amplitudes. Therefore they are unlikely to be the
HKLM ground state. Ruling out these six $Z_2$ SLs, we can see the
other 14 $Z_2$ SL states fall into 4 classes. To be specific, they
are continuously connected to different parent $U(1)$ gapless SL states on
$\kag$ lattice. These parent $U(1)$ SL states in general have the following
mean-field ansatz
\begin{eqnarray}\label{MF:U1}
H_{U(1)SL}=\chi_1\sum_{<ij>}\nu_{ij}(f^\dagger_{i\alpha}f_{j\alpha}+~h.c.)
\end{eqnarray}
where $\nu_{ij}=\pm1$ characterizes the sign structure of hopping
terms with $\chi_1\in\mathbb{R}$. Different parent $U(1)$ SL states
are featured by the flux of $f$-spinon hopping phases around basic
plaquette: honeycombs and triangles on the $\kag$ lattice.

The simplest example is the so-called uniform RVB state with
$\nu_{ij}\equiv+1$ for all 1st n.n. mean-field bonds. The hopping
phase around any plaquette is $1=\exp[\imth 0]$, and the
corresponding flux is $[0,0]$ for [triangle,honeycomb] motifs. The 4
possible $Z_2$ spin liquids in the neighborhood\cite{Lu2010} of
uniform RVB states (\ie $U(1)$ SL-$[0,0]$ state in \Ref{Ran2007})
are classified in Appendix \ref{app:Z2 around u-RVB}. They are
$\#1,\#5,\#15,\#13$ in TABLE \ref{tab:MF_ansatz} and TABLE
\ref{tab:z2kagome}. We label them as $Z_2[0,0]A$, $Z_2[0,0]B$,
$Z_2[0,0]C$ and $Z_2[0,0]D$ states. They all have gapped spectra of
spinons.

\begin{figure}
 \includegraphics[width=0.22\textwidth]{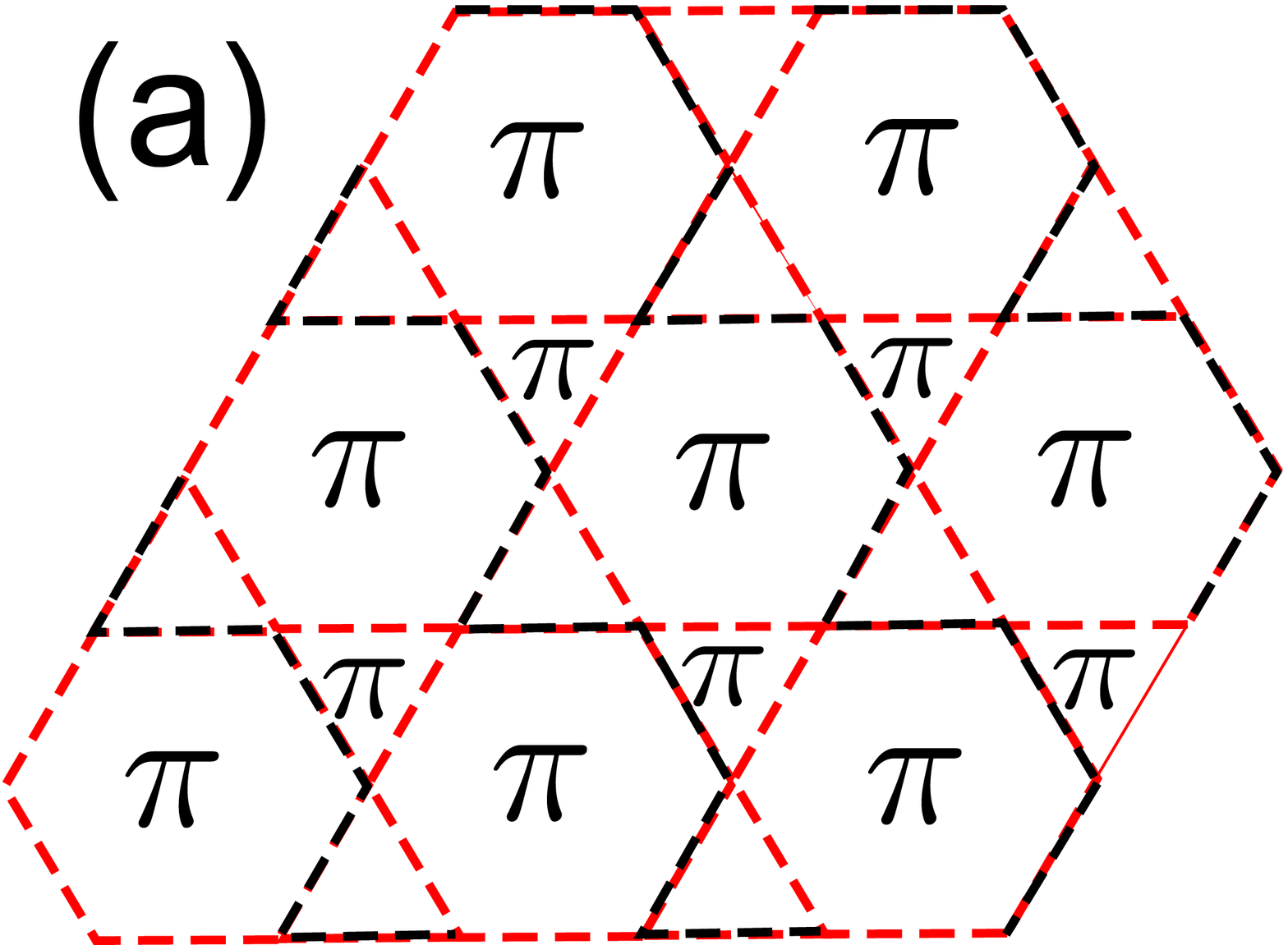}\;\includegraphics[width=0.22\textwidth]{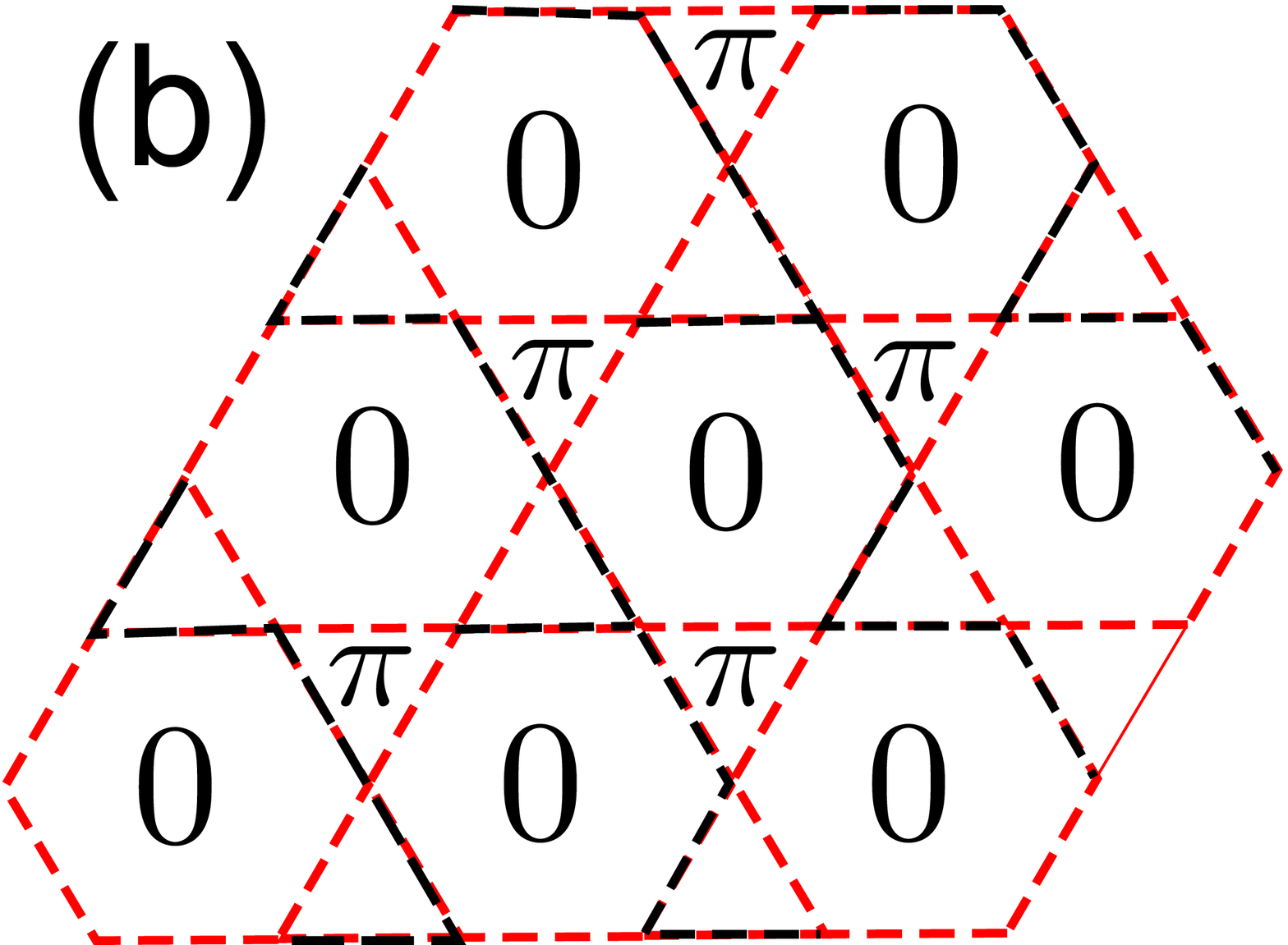}
\caption{(color online) Mean-field ansatz of (a) $U(1)$
SL-$[\pi,\pi]$ state and (b) $U(1)$ SL-$[\pi,0]$ state, with 1st
n.n. real hopping terms
$H_{MF}=\chi_1\sum_{<ij>\alpha}(\nu_{ij}f^\dagger_{i\alpha}f_{j\alpha}+~h.c.)$.~Colors
again denote the sign structure of mean-field bonds: red dashed
lines have $\nu_{ij}=+1$ and black dashed lines have $\nu_{ij}=-1$.}
\label{fig:U1-SL-[pi,0/pi]}
\end{figure}

The ansatz of two other parent $U(1)$ SLs are shown in FIG.
\ref{fig:U1-SL-[pi,0/pi]}. They both have $\pi$-flux piercing
through a triangle basic plaquette. Following the above notations of
hopping phase in [triangle,honeycomb] motifs, with either $\pi$-flux
or $0$-flux through the honeycomb plaquette, they are called $U(1)$
SL-$[\pi,\pi]$ state and $U(1)$ SL-$[\pi,0]$ state. There are three
$Z_2$ SLs in the neighborhood of both $U(1)$ SL states, \ie
$\#3,\#17,\#19$ around $U(1)$ SL-$[\pi,\pi]$ state and
$\#4,\#18,\#20$ around $U(1)$ SL-$[\pi,0]$ state. All these six
$Z_2$ SLs have gapless spinon spectra, inherited from the two parent gapless $U(1)$ SLs. To be precise, the spinon band structure of these six $Z_2$ SL states are featured by a doubly-degenerate flat band and a
Dirac cone at Brillouin-zone center. This is in contrast to the
numerically observed gap in two-spinon spectrum\cite{Yan2010}, thus
we can also rule out these 6 $Z_2$ SLs for the HKLM ground state.

Another $U(1)$ SL state is the so called $U(1)$-Dirac SL or $U(1)$
SL-$[0,\pi]$ state. Its mean-field ansatz is shown by the 1st n.n.
bonds in FIG. \ref{fig:kagome}(b). Clearly $\pi$-flux pierces
through certain triangle plaquette with no flux through the
honeycomb plaquette. According to variational Monte Carlo
studies\cite{Hastings2000,Ran2007}, this $U(1)$-Dirac SL have
substantially lower energy compared to many other competing phases,
including the uniform RVB state. Therefore we shall focus on
those $Z_2$ SLs in the neighborhood of the $U(1)$-Dirac SL in our
search of the HKLM ground state. We need to mention that although
unlikely, the four $Z_2$ SLs in the neighborhood of uniform RVB
state, or $U(1)$ SL-$[0,0]$ state are potentially possible to be the
HKLM ground state.

In a previous study using PSG in Schwinger-boson representation\cite{Wang2006}, it was shown that there are 8 different Schwinger-boson mean-field ansatz of $Z_2$ SLs on the $\kag$ lattice which preserve all lattice symmetry. However, these 8 $Z_2$ SLs may or may not preserve time-reversal symmetry. One can show that requiring all lattice symmetry \emph{and time-reversal symmetry}, there are 16 different Schwinger-boson $Z_2$ SLs on the $\kag$ lattice. The relation between the 20 $Z_2$ SLs in Schwinger-fermion representation (see TABLE \ref{tab:MF_ansatz}) and the 16 $Z_2$ SLs in Schwinger-boson representation are not clear.
To clarify the relation between SL states in these two different representations, one can compare the neighboring (ordered) phases of the SLs, \eg by computing the vison quantum numbers\cite{Lu2010a} of SL states.


\subsection{$Z_2[0,\pi]\beta$ state as a promising candidate for the HKLM ground state}

How to find those $Z_2$ SLs in the neighborhood of (or continuously connected to) the $U(1)$-Dirac SL? Naively, we expect the mean-field
ansatz of these $Z_2$ SLs can be obtained from that of $U(1)$-Dirac
SL by adding an infinitesimal perturbation. To be specific, we
require an infinitesimal spinon pairing term on top of the
$U(1)$-Dirac SL mean-field ansatz (\ref{MF:U1}) or (\ref{MF:U1SL0})
to break the IGG from $U(1)$ to $Z_2$ through Higgs mechanism.
Mathematically, we need to find those $Z_2$ SL states whose PSG is a
subgroup of the $U(1)$-Dirac SL's PSG\cite{Lu2010}. Such $Z_2$ SL
states are defined to be in the neighborhood of $U(1)$-Dirac SL.
Similar criterion applies to the neighboring $Z_2$ SL states of any parent
$U(1)$ or $SU(2)$ SL state.

We find out all four $Z_2$ SLs in the neighborhood of $U(1)$-Dirac
SLs in Appendix \ref{app:Z2 around U1 Dirac}. They are states
$\#6,\#2,\#14,\#16$ in TABLE \ref{tab:MF_ansatz}, labeled as
$Z_2[0,\pi]\alpha$, $Z_2[0,\pi]\beta$, $Z_2[0,\pi]\gamma$ and
$Z_2[0,\pi]\delta$ states respectively. Since the effective theory
of a $U(1)$-Dirac SL is an 8-component Dirac fermion coupled with
dynamical $U(1)$ gauge field\cite{Ran2007,Hermele2008}, we can find
out all symmetry-allowed mass terms that can open up a gap in the
Dirac-like spinon spectrum. Following detailed calculations in
Appendix \ref{app:Z2 around U1 Dirac}, we can see that among the
four $Z_2$ SLs around the $U(1)$-Dirac SL, only one state, \ie
$Z_2[0,\pi]\beta$ (state $\#2$ in TABLE \ref{tab:MF_ansatz} and
\ref{tab:z2kagome}) can generate a mass gap in the spinon spectrum. In other 3 states the Dirac cone in spinon spectrum is
protected by symmetry. The mean-field ansatz of $Z_2[0,\pi]\beta$ SL
state up to 2nd n.n. is shown in FIG. \ref{fig:kagome}(b):
\begin{eqnarray}\label{mf:z2[0,pi]beta}
&H_{MF}=\sum_{i}(\lambda_3\sum_\alpha f^\dagger_{i\alpha}f_{i\alpha}+\lambda_1f^\dagger_{i\uparrow}f^\dagger_{i\downarrow}+h.c.)\\
&\notag+\chi_1\sum_{<ij>\alpha}\nu_{ij}(f^\dagger_{i\alpha}f_{j\alpha}+h.c.)+\\
&\notag\sum_{<<ij>>}\nu_{ij}(\chi_2\sum_\alpha
f^\dagger_{i\alpha}f_{j\alpha}+\Delta_2\sum_{\alpha\beta}\epsilon^{\alpha\beta}f^\dagger_{i\alpha}f^\dagger_{j\beta}+h.c.)
\end{eqnarray}
where $\epsilon^{\alpha\beta}$ is the completely anti-symmetric tensor. We only
list up to 2nd n.n. mean-field amplitudes because as shown in TABLE
\ref{tab:MF_ansatz} (see also Appendix \ref{app:Z2 around U1
Dirac}), this $Z_2[0,\pi]\beta$ state only needs 2nd n.n. pairing
terms to realize a $Z_2$ SL. We can always choose a proper gauge so
that mean-field parameters $\chi_{1,2}$ and $\Delta_2$ are all real.
The sign structure of $\nu_{ij}=\pm1$ are shown in FIG.
\ref{fig:kagome}(b), with red denoting $\nu_{ij}=+1$ and other colors
representing $\nu_{ij}=-1$. As discussed in Appendix \ref{app:Z2
around U1 Dirac}, the 2nd n.n. singlet-pairing term $\Delta_2\neq0$
not only break the $U(1)$ gauge symmetry down to $Z_2$, but also
opens up a mass gap in the spinon spectrum. The on-site chemical
potential $\lambda_{1,3}$ are self-consitently determined by the
following constraint:
\begin{eqnarray}\label{constraint:mf}
&\notag\sum_i\langle
f^\dagger_{i\uparrow}f^\dagger_{i\downarrow}\rangle=\sum_i\langle
f_{i\uparrow}f_{i\downarrow}\rangle=0,\\
&\sum_i(\sum_{\alpha=\uparrow,\downarrow}f^\dagger_{i\alpha}f_{i\alpha}-1)=0.
\end{eqnarray}
For further n.n. mean-field ansatz see discussions in Appendix
\ref{app:Z2 around U1 Dirac}.

\section{Conclusion}

To summarize, motivated by the strong evidence of a $Z_2$ SL as the
HKLM ground state in recent DMRG study\cite{Yan2010}, we classify
all possible $Z_2$ SL states in Schwinger-fermion mean-field
approach using PSG. We found 20 different Schwinger-fermion
mean-field states of $Z_2$ SLs on $\kag$ lattice, among which 6
states are unlikely due to vanishing 1st n.n. mean-field amplitude.
In other 14 $Z_2$ SLs only 5 possess a gapped spinon spectrum, which
is observed in the DMRG result\cite{Yan2010}. These five
symmetric $Z_2$ SL states are all in the neighborhood of certain parent $U(1)$ gapless SLs. To be precise, four are in the neighborhood of gapless uniform
RVB (or $U(1)$ SL-$[0,0]$) state, while the other one, \ie
$Z_2[0,\pi]\beta$ is in the neighborhood of gapless $U(1)$-Dirac SL (or $U(1)$ SL-$[0,\pi]$) state.
Previous variational Monte Carlo study\cite{Ran2007} showed that
gapless $U(1)$-Dirac SL has a substantially lower energy in comparison to
the uniform RVB state. This suggests $Z_2$ SLs in the neighborhood of $U(1)$-Dirac SL should have lower energy compared to those in the neighborhood of uniform RVB state.
Therefore we propose this $Z_2[0,\pi]\beta$
state with mean-field ansatz (\ref{mf:z2[0,pi]beta}) shown in FIG.
\ref{fig:kagome}(b) as the HKLM ground state numerically detected in
\Ref{Yan2010}. Our work provides important insight for future
numeric study, e.g. variational Monte Carlo study of Gutzwiller projected wavefunctions.

\begin{acknowledgements}

YML thank Prof. Ziqiang Wang for support under DOE Grant
DE-FG02-99ER45747. YR is supported by the startup fund at Boston
College. PAL acknowledges the support under NSF DMR-0804040.

\end{acknowledgements}

\appendix

\section{Symmetry group of kagome lattice and algebra conditions for $Z_2$ spin
liquids}\label{app:symmetry group}

As shown in FIG. \ref{fig:kagome}(a), we label the three lattice
sites in each unit cell with sublattice index $\{s=u,v,w\}$.
Choosing Bravais unit vector as $\vec a_1=a\hat x$ and $\vec
a_2=\frac{a}2(\hat x+\sqrt3\hat y)$, the positions of the three
atoms in a unit cell labeled by indices $i=(x,y,s)$ are
\begin{eqnarray}
&\vec r(x,y,u)=(x+\frac12)\vec a_1+(y+\frac12)\vec a_2,\\
&\notag\vec r(x,y,v)=(x+\frac12)\vec a_1+y\vec a_2,\\
&\notag\vec r(x,y,w)=x\vec a_1+(y+\frac12)\vec a_2.
\end{eqnarray}
The symmetry group of such a two-dimensional $\kag$ lattice is
generated by the following operations
\begin{eqnarray}
&T_1:~~~(x,y,s)\rightarrow(x+1,y,s);\\
&\notag T_2:~~~(x,y,s)\rightarrow(x,y+1,s);\\
&\notag \bss
:~~~(x,y,u)\rightarrow (y,x,u),\\
&\notag ~~~(x,y,v)\rightarrow(y,x,w),\\
&\notag ~~~(x,y,w)\rightarrow(y,x,v);\\
&\notag \cs:~~~(x,y,u)\rightarrow(-y-1,x+y+1,v),\\
&\notag ~~~(x,y,v)\rightarrow(-y,x+y,w),\\
&\notag ~~~(x,y,w)\rightarrow(-y-1,x+y,u).
\end{eqnarray}
together with time reversal $\bst$.

The symmetry group of a $\kag$ lattice is defined by the following
algebraic relations between its generators:
\begin{eqnarray}\label{algebra:sg}
&\bst^2=\bss^2=(\cs)^6=\bse,\\
&\notag g^{-1}\bst^{-1}g\bst=\bse,~~~\forall~g=T_{1,2},\bss,\cs,\\
&\notag T_2^{-1}T_1^{-1}T_2T_1=\bse,\\
&\notag \bss^{-1}T_1^{-1}\bss T_2=\bse,\\
&\notag \bss^{-1}T_2^{-1}\bss T_1=\bse,\\
&\notag \cs^{-1}T_2^{-1}\cs T_1=\bse,\\
&\notag \cs^{-1}T_2^{-1}T_1\cs T_2=\bse,\\
&\notag \bss^{-1}\cs\bss\cs=\bse.
\end{eqnarray}
where $\bse$ stands for the identity element in the symmetry group.
Therefore the consistent conditions for a generic $Z_2$ PSGs on a
$\kag$ lattice is written as
\begin{eqnarray}\label{algebra:psg:T}
&[G_\bst(i)]^2=\eta_T\tau^0,\\
&G_\bss(\bss(i))G_\bss(i)=\eta_\bss\tau^0,\label{algebra:psg:sig}\\
&
G^\dagger_{T_1}(i)G_\bst^\dagger(i)G_{T_1}(i)G_\bst({T_1}^{-1}(i))=\eta_{T_1\bst}\tau^0,\label{algebra:psg:T1,T}\\
&
G^\dagger_{T_2}(i)G_\bst^\dagger(i)G_{T_2}(i)G_\bst({T_2}^{-1}(i))=\eta_{T_2\bst}\tau^0,\label{algebra:psg:T2,T}\\
&
G^\dagger_{\bss}(i)G_\bst^\dagger(i)G_\bss(i)G_\bst(\bss^{-1}(i))=\eta_{\bss\bst}\tau^0,\label{algebra:psg:sig,T}\\
&
G^\dagger_\cs(i)G_\bst^\dagger(i)G_\cs(i)G_\bst(\cs^{-1}(i))=\eta_{\cs\bst}\tau^0,\label{algebra:psg:C6,T}\\
&
G^\dagger_{T_2}(T_1^{-1}(i))G^\dagger_{T_1}(i)G_{T_2}(i)G_{T_1}(T_2^{-1}(i))=\eta_{12}\tau^0,\label{algebra:psg:T1,T2}\\
&\notag
G_\cs(\cs^{-1}(i))G_\cs(\cs^{-2}(i))G_\cs(\cs^3(i))G_\cs(\cs^2(i))\cdot\\
&G_\cs(\cs^2(i))G_\cs(\cs(i))G_\cs(i)=\eta_\cs\tau^0,\label{algebra:psg:C6}\\
&
G^\dagger_\bss(T_2^{-1}(i))G^\dagger_{T_2}(i)G_\bss(i)G_{T_1}(\bss(i))=\eta_{\bss
T_1}\tau^0,\label{algebra:psg:sig,T1}\\
&
G^\dagger_\bss(T_1^{-1}(i))G^\dagger_{T_1}(i)G_\bss(i)G_{T_2}(\bss(i))=\eta_{\bss
T_2}\tau^0,\label{algebra:psg:sig,T2}\\
&
G^\dagger_\bss(\cs(i))G_\cs(\cs(i))G_\bss(i)G_\cs(\bss(i))=\eta_{\bss\cs}\tau^0,\label{algebra:psg:sig,C6}\\
&
G^\dagger_\cs(T_2^{-1}(i))G^\dagger_{T_2}(i)G_\cs(i)G_{T_1}(\cs^{-1}(i))=\eta_{\cs T_1}\tau^0,\label{algebra:psg:C6,T1}\\
&\notag
G^\dagger_\cs(T_2^{-1}T_1(i))G^\dagger_{T_2}(T_1(i))G_{T_1}(T_1(i))\cdot\\
& G_\cs(i)G_{T_2}(\cs^{-1}(i))=\eta_{\cs
T_2}\tau^0.\label{algebra:psg:C6,T2}
\end{eqnarray}
for any lattice site $i=(x,y,s)$. Here all
$\eta$s~
are $Z_2$ integers characterizing different SLs: different (gauge
inequivalent) choices of these $Z_2$ integers (different $Z_2$ PSGs)
correspond to different $Z_2$ SLs. Notice that under a local guage
transformation $W(i)\in SU(2)$ the PSG element $G_U(i)$ transforms
as
\begin{eqnarray}
G_U(i)\rightarrow W(i)G_U(i)W^\dagger(U^{-1}(i))
\end{eqnarray}

\section{Classification of all $Z_2$ spin liquids on kagome
lattice}

\subsection{Classification of $Z_2$ algebraic PSGs on $\kag$
lattice}\label{app:kagome Z2 PSGs}

In this section we classify all possible $Z_2$ spin liquids on a
$\kag$ lattice. Mathematically we need to find out all
gauge-inequivalent solutions of algebraic conditions
(\ref{algebra:psg:T})-(\ref{algebra:psg:C6,T1}) for $Z_2$ PSGs.

First from condition (\ref{algebra:psg:T1,T2}) we can always choose
a proper gauge so that
\begin{eqnarray}\label{psg:T1,2}
G_{T_1}(x,y,s)=\eta_{12}^y\tau^0,~~~G_{T_2}(x,y,s)\equiv\tau^0.
\end{eqnarray}

From (\ref{algebra:psg:sig,T1}) and (\ref{algebra:psg:sig,T2}) we
can see $G_\bss(x,y,s)=\eta_{\bss T_1}^y\eta_{\bss
T_2}^x\eta_{12}^{xy}g_\bss(s)$. Condition (\ref{algebra:psg:sig})
further determine \alert{$\eta_{\bss T_1}=\eta_{\bss T_2}$} and
therefore we have
\begin{eqnarray}\notag
G_\bss(x,y,s)=\eta_{\bss T_1}^{x+y}\eta_{12}^{xy}g_\bss(s)
\end{eqnarray}
where $SU(2)$ matrices $g_\bss(s)$ satisfy
\begin{eqnarray}
g_\bss(w)g_\bss(v)=\big[g_\bss(u)\big]^2=\eta_\bss\tau^0\label{condition:sig}
\end{eqnarray}

Notice that we can always choose a proper global $Z_2$ gauge on
$G_{T_1}(x,y,s)$ (which doesn't change the mean-field ansatz) so
that \alert{$\eta_{\cs T_2}=1$} in (\ref{algebra:psg:C6,T2}). From
(\ref{algebra:psg:C6,T1}) and (\ref{algebra:psg:C6,T2}) it's
straightforward to show that $G_\cs(x,y,u/v)=\eta_{\cs
T_1}^{x+y}\eta_{12}^{xy+x(x+1)/2}g_\cs(u/v)$ and
$G_\cs(x,y,w)=\eta_{\cs
T_1}^{x+y}\eta_{12}^{x+y+xy+x(x+1)/2}g_\cs(w)$. It's condition
(\ref{algebra:psg:sig,C6}) that determines \alert{$\eta_{\cs
T_1}=\eta_{\bss T_1}\eta_{12}$} and finally we have
\begin{eqnarray}
&\notag G_\cs(x,y,u/v)=\eta_{\bss T_1}^{x+y}\eta_{12}^{xy+\frac{x(x+1)}2}g_\cs(u/v),\\
&G_\cs(x,y,w)=(\eta_{12}\eta_{\bss
T_1})^{x+y}\eta_{12}^{xy+\frac{x(x+1)}2}g_\cs(w).\notag
\end{eqnarray}
where $SU(2)$ matrices $g_\cs(s)$ satisfy
\begin{eqnarray}
&\label{condition:C6}\big[g_\cs(w)g_\cs(v)g_\cs(u)\big]^2=\eta_{12}\eta_\cs\tau^0,\\
&\notag\big[g_\bss(v)g_\cs(w)\big]^2=g_\bss(w)g_\cs(v)g_\bss(u)g_\cs(u)=\eta_\bss\eta_{\bss\cs}\tau^0.\\
\label{condition:C6_sig}
\end{eqnarray}
according to (\ref{algebra:psg:C6}) and (\ref{algebra:psg:sig,C6}).

Now through a gauge transformation $W(x,y,s)=\eta_{\bss T_1}^y$ we
can fix \alert{$\eta_{\bss T_{1,2}}=1$} and the PSG elements become
\begin{eqnarray}
&\label{psg:sig} G_\bss(x,y,s)=\eta_{12}^{xy}g_\bss(s);\\
&\notag G_\cs(x,y,u/v)=\eta_{12}^{xy+\frac{x(x+1)}2}g_\cs(u/v),\\
&G_\cs(x,y,w)=\eta_{12}^{xy+x+y+\frac{x(x+1)}2}g_\cs(w).\label{psg:C6}
\end{eqnarray}

According to (\ref{algebra:psg:T}), (\ref{algebra:psg:T1,T}) and
(\ref{algebra:psg:T2,T}) we can see that
$G_\bst(x,y,s)=\eta^x_{T_1\bst}\eta^y_{T_2\bst}g_\bst(s)$.
(\ref{algebra:psg:C6,T}) and (\ref{algebra:psg:sig,T}) further
determines \alert{$\eta_{T_1\bst}=\eta_{T_2\bst}=1$} and by choosing
a proper gauge we have
\begin{eqnarray}
G_\bst(x,y,s)=g_\bst(s)\equiv\left.\Big\{\begin{aligned}\tau^0,~~~&\eta_\bst=1.\\\imth\tau^1,~~~&\eta_\bst=-1.\end{aligned}\right.
\end{eqnarray}
which satisfy
\begin{eqnarray}
&g_\bss(u)g_\bst(u)=\eta_{\bss\bst}g_\bst(u)g_\bss(u),\label{condition:T_sig}\\
&\notag g_\bss(v)g_\bst(w)=\eta_{\bss\bst}g_\bst(v)g_\bss(v),\\
&\notag g_\bss(w)g_\bst(v)=\eta_{\bss\bst}g_\bst(w)g_\bss(w);\\
&g_\cs(u)g_\bst(w)=\eta_{\cs\bst}g_\bst(u)g_\cs(u),\label{condition:T_C6}\\
&\notag g_\cs(v)g_\bst(u)=\eta_{\cs\bst}g_\bst(v)g_\cs(v),\\
&\notag g_\cs(w)g_\bst(v)=\eta_{\cs\bst}g_\bst(w)g_\cs(w).
\end{eqnarray}
according to (\ref{algebra:psg:C6,T}) and (\ref{algebra:psg:sig,T}).

\begin{table}[tb]
\begin{tabular}{|c||c|c|c|c|c|c|c|c|}
\hline $\#$ & $\eta_{12}$&$g_{\bss}(u)$ & $g_\bss(v)$&$g_\bss(w)$ & $g_\cs(u)$& $g_\cs(v)$&$g_\cs(w)$&Label\\
\hline 1&$+1$&$\tau^0$&$\tau^0$&$\tau^0$&$\tau^0$&$\tau^0$&$\tau^0$&$Z_2[0,0]A$\\
\hline 2&$-1$&$\tau^0$&$\tau^0$&$\tau^0$&$\tau^0$&$\tau^0$&$\tau^0$&$Z_2[0,\pi]\beta$\\
\hline 3&$+1$&$\tau^0$&$\tau^0$&$\tau^0$&$\tau^0$&$-\tau^0$&$\imth\tau^1$&$Z_2[\pi,\pi]A$\\
\hline 4&$-1$&$\tau^0$&$\tau^0$&$\tau^0$&$\tau^0$&$-\tau^0$&$\imth\tau^1$&$Z_2[\pi,0]A$\\
\hline 5&$+1$&$\tau^0$&$\tau^0$&$\tau^0$&$\imth\tau^3$&$\imth\tau^3$&$\imth\tau^3$&$Z_2[0,0]B$\\
\hline 6&$-1$&$\tau^0$&$\tau^0$&$\tau^0$&$\imth\tau^3$&$\imth\tau^3$&$\imth\tau^3$&$Z_2[0,\pi]\alpha$\\
\hline 7&$+1$&$\imth\tau^1$&$\tau^0$&$-\tau^0$&$\tau^0$&$\imth\tau^1$&$\tau^0$&-\\
\hline 8&$-1$&$\imth\tau^1$&$\tau^0$&$-\tau^0$&$\tau^0$&$\imth\tau^1$&$\tau^0$&-\\
\hline 9&$+1$&$\imth\tau^1$&$\tau^0$&$-\tau^0$&$\tau^0$&$-\imth\tau^1$&$\imth\tau^1$&-\\
\hline 10&$-1$&$\imth\tau^1$&$\tau^0$&$-\tau^0$&$\tau^0$&$-\imth\tau^1$&$\imth\tau^1$&-\\
\hline 11&$+1$&$\imth\tau^1$&$\tau^0$&$-\tau^0$&$\imth\tau^3$&$-\imth\tau^2$&$\imth\tau^3$&-\\
\hline 12&$-1$&$\imth\tau^1$&$\tau^0$&$-\tau^0$&$\imth\tau^3$&$-\imth\tau^2$&$\imth\tau^3$&-\\
\hline 13&$+1$&$\imth\tau^3$&$\imth\tau^3$&$\imth\tau^3$&$\imth\tau^3$&$\imth\tau^3$&$\imth\tau^3$&$Z_2[0,0]D$\\
\hline 14&$-1$&$\imth\tau^3$&$\imth\tau^3$&$\imth\tau^3$&$\imth\tau^3$&$\imth\tau^3$&$\imth\tau^3$&$Z_2[0,\pi]\gamma$\\
\hline 15&$+1$&$\imth\tau^3$&$\imth\tau^3$&$\imth\tau^3$&$\tau^0$&$\tau^0$&$\tau^0$&$Z_2[0,0]C$\\
\hline 16&$-1$&$\imth\tau^3$&$\imth\tau^3$&$\imth\tau^3$&$\tau^0$&$\tau^0$&$\tau^0$&$Z_2[0,\pi]\delta$\\
\hline 17&$+1$&$\imth\tau^3$&$\imth\tau^3$&$\imth\tau^3$&$\tau^0$&$\tau^0$&$\imth\tau^1$&$Z_2[\pi,\pi]B$\\
\hline 18&$-1$&$\imth\tau^3$&$\imth\tau^3$&$\imth\tau^3$&$\tau^0$&$\tau^0$&$\imth\tau^1$&$Z_2[\pi,0]B$\\
\hline 19&$+1$&$\imth\tau^3$&$\imth\tau^3$&$\imth\tau^3$&$\imth\tau^3$&$-\imth\tau^3$&$\imth\tau^2$&$Z_2[\pi,\pi]C$\\
\hline 20&$-1$&$\imth\tau^3$&$\imth\tau^3$&$\imth\tau^3$&$\imth\tau^3$&$-\imth\tau^3$&$\imth\tau^2$&$Z_2[\pi,0]C$\\
\hline
\end{tabular}
\caption{\label{tab:z2kagome}A summary of all 20 gauge-inequivalent
PSG's with $G_\bst(x,y,s)=\imth\tau^1$ on the $\kag$ lattice. Notice
that there is a free $Z_2$ integer $\eta_{12}=\pm1$ in other PSG
elements (\ref{psg:T1,2}), (\ref{psg:sig}) and (\ref{psg:C6}). They
correspond to 20 different $Z_2$ spin liquids on the $\kag$
lattice.}
\end{table}

In the following we find out all the gauge-inequivalent solutions of
$SU(2)$ matrices $g_{\bst,\bss,\cs}(s)$ satisfying the above
conditions. They are summarized in TABLE .

(\Rmnum{1}) $g_\bst(s)=\tau^0$ and therefore
\alert{$\eta_\bst=\eta_{\bss\bst}=\eta_{\cs\bst}=1$}:

Conditions (\ref{condition:T_sig}) and (\ref{condition:T_C6}) are
automatically satisfied.

(\rmnum{1}) \alert{$\eta_\bss=1$}:

Notice that under a global gauge transformation $W(x,y,s)\equiv
W_s\in SU(2)$ the PSG elements transform as
\begin{eqnarray}
&\notag g_\bss(u)\rightarrow W_ug_\bss(u)W^\dagger_u,\\
&\notag g_\bss(v)\rightarrow W_vg_\bss(v)W^\dagger_w,\\
&\notag g_\bss(w)\rightarrow W_wg_\bss(w)W^\dagger_v;\\
&\notag g_\cs(u)\rightarrow W_ug_\cs(u)W^\dagger_w,\\
&\notag g_\cs(v)\rightarrow W_vg_\cs(v)W^\dagger_u,\\
&\notag g_\cs(w)\rightarrow W_wg_\cs(w)W^\dagger_v.
\end{eqnarray}
Thus from (\ref{condition:sig}) and (\ref{condition:C6_sig}) we can
always have $g_\bss(s)=\tau^0$ and $g_\cs(u)=\tau^0$,
$g_\cs(v)=\eta_{\bss\cs}\tau^0$ by choosing a proper gauge.

(A) \alert{$\eta_{\bss\cs}=\eta_{12}\eta_{\cs}=1$}:

from (\ref{condition:C6}) we have $g_\cs(w)=\tau^0$.

(B) \alert{$\eta_{\bss\cs}=\eta_{12}\eta_{\cs}=-1$}:

from (\ref{condition:C6}) we have $g_\cs(w)=\imth\tau^3$ by gauge
fixing.\\

(\rmnum{2}) \alert{$\eta_\bss=-1$}:

from (\ref{condition:sig}) we have $g_\bss(v)=-g_\bss(w)=\tau^0$ and
$g_\bss(u)=\imth\tau^3$ by gauge fixing. Also from
(\ref{condition:C6_sig}) we can choose a gauge so that
$g_\cs(u)=\tau^0$ and $g_\cs(v)=-\imth\eta_{\bss\cs}\tau^3$.

(A) \alert{$\eta_{\bss\cs}=-1$}:

In this case (\ref{condition:C6_sig}) requires $g_\cs(w)=\tau^0$ and
thus \alert{$\eta_{12}\eta_\cs=-1$} according to
(\ref{condition:C6}).

(B) \alert{$\eta_{\bss\cs}=1$}:

(a) \alert{$\eta_{12}\eta_{\cs}=-1$}:

Now from (\ref{condition:C6_sig}) and (\ref{condition:C6}) we have
$g_\cs(w)=\imth\tau^1$ by gauge fixing.

(b) \alert{$\eta_{12}\eta_{\cs}=1$}:

by (\ref{condition:C6_sig}) and (\ref{condition:C6}) we must have
$g_\cs(w)=\imth\tau^3$.

To summarize there are $2\times(2+3)=10$ different algebraic PSGs
with
$\eta_\bst=1$ and $g_\bst(s)=\tau^0$.\\

(\Rmnum{2}) $g_\bst(s)=\imth\tau^1$ and \alert{$\eta_\bst=-1$}:

(\rmnum{1}) \alert{$\eta_\bss=1$}:

According to (\ref{condition:sig})and (\ref{condition:T_sig}), by
choosing a proper gauge we can have $g_\bss(s)=\tau^0$ and
\alert{$\eta_{\bss\bst}=1$}. From (\ref{condition:C6}) and
(\ref{condition:C6_sig}) we also have
$\big[g_\cs(w)\big]^2=g_\cs(v)g_\cs(u)=\eta_{\bss\cs}\tau^0=\eta_{12}\eta_\cs\tau^0$.

(A) \alert{$\eta_{12}\eta_\cs=\eta_{\bss\cs}=1$}:

From (\ref{condition:T_C6}), (\ref{condition:C6}) and
(\ref{condition:C6_sig}), by choosing gauge we have
$g_\cs(s)=\tau^0$ and \alert{$\eta_{\cs\bst}=1$}.

(B) \alert{$\eta_{12}\eta_\cs=\eta_{\bss\cs}=-1$}:

(a) \alert{$\eta_{\cs\bst}=1$}:

In this case we have $g_\cs(u)=-g_\cs(v)=\tau^0$ and
$g_\cs(w)=\imth\tau^1$ by choosing a proper gauge.

(b) \alert{$\eta_{\cs\bst}=-1$}:

In this case we can have $g_\cs(s)=\imth\tau^3$ by choosing a proper gauge.\\

(\rmnum{2}) \alert{$\eta_\bss=-1$}:

(A) \alert{$\eta_{\bss\bst}=1$}:

From (\ref{condition:T_sig}) and (\ref{condition:sig}) we have
$g_\bss(u)=\imth\tau^1$ and $g_\bss(v)=-g_\bss(w)=\tau^0$ by proper
gauge fixing. Also from (\ref{condition:C6_sig}) we know
$\big[g_\cs(w)\big]^2=-\eta_{\bss\cs}\tau^0$ and
$g_\cs(u)g_\cs(v)=-\imth\eta_{\bss\cs}\tau^1$.

(a) \alert{$\eta_{\bss\cs}=-1$}:

from (\ref{condition:T_C6}) and (\ref{condition:C6_sig}),
(\ref{condition:C6}) it's clear that \alert{$\eta_{\cs\bst}=1$},
$g_\cs(u)=g_\cs(w)=\tau^0$ and $g_\cs(v)=\imth\tau^1$ through gauge
fixing. Also we have \alert{$\eta_{12}\eta_\cs=-1$}.

(b) \alert{$\eta_{\bss\cs}=1$}:

(b1) \alert{$\eta_{\cs\bst}=1$}:

In this case \alert{$\eta_{12}\eta_{\cs}=1$}, and we can always
choose a proper gauge so that
$g_\cs(u)=\tau^0$,~$g_\cs(w)=-g_\cs(v)=\imth\tau^1$.

(b2) \alert{$\eta_{\cs\bst}=-1$}:

In this case \alert{$\eta_{12}\eta_{\cs}=-1$}, and we can always
choose a proper gauge so that
$g_\cs(v)=-\imth\tau^2$,~$g_\cs(u)=g_\cs(w)=\imth\tau^3$.

(B) \alert{$\eta_{\bss\bst}=-1$}:

Conditions (\ref{condition:T_sig}) and (\ref{condition:sig}) assert
that $g_\bss(s)=\imth\tau^3$ by proper gauge choosing.

(a) \alert{$\eta_{\bss\cs}=-1$}:

In this case from (\ref{condition:C6_sig}) we know
$g_\cs(w)=\imth\tau^3$, hence \alert{$\eta_{\cs\bst}=-1$}. Then we
can always choose a gauge so that $g_\cs(u)=g_\cs(v)=\imth\tau^3$
and so \alert{$\eta_{12}\eta_\cs=-1$} from (\ref{condition:C6}).

(b) \alert{$\eta_{\bss\cs}=1$}:

(b1) \alert{$\eta_{\cs\bst}=1$}:

In this case from (\ref{condition:T_sig}),(\ref{condition:C6_sig})
we have $g_\cs(u)=g_\cs(v)=\tau^0$ by a proper gauge choice.
Meanwhile conditions (\ref{condition:C6}) and
(\ref{condition:C6_sig}) become
$\big[g_\cs(w)\big]^2=\eta_{12}\eta_{\cs}\tau^0$ and
$\big[\imth\tau^3g_\cs(w)\big]^2=-\tau^0$.

(b.1.1) \alert{$\eta_{12}\eta_\cs=1$}:

here we have $g_\cs(w)=\tau^0$.

(b.1.2) \alert{$\eta_{12}\eta_\cs=-1$}:

here we have $g_\cs(w)=\imth\tau^1$.

(b2) \alert{$\eta_{\cs\bst}=-1$}:

In this case from (\ref{condition:T_sig}) and
(\ref{condition:C6_sig}) we can always choose a proper gauge so that
$g_\cs(u)=-g_\cs(v)=\imth\tau^3$. We also have
$g_\cs(w)=\imth\tau^2$ and \alert{$\eta_{12}\eta_\cs=-1$} from
(\ref{condition:C6}).

To summarize there are $2\times(3+7)=20$ different algebraic PSGs
with
$\eta_\bst=-1$ and $g_\bst(s)=\imth\tau^1$.\\

So in summary we have $10+20=30$ different $Z_2$ algebraic PSGs
satisfying conditions
(\ref{algebra:psg:T})-(\ref{algebra:psg:C6,T2}). Among them there
are at most 20 solutions that can be realized by a mean-field
ansatz, since those PSGs with $g_\bst(s)=\tau^0$ would require all
mean-field bonds to vanish due to (\ref{psg:def:T}). As a result
there are \alert{20} different $Z_2$ spin liquids on a $\kag$
lattice.

\subsection{Symmetry conditions on mean-field
anstaz}\label{app:MF_cond}

Let's denote the mean-field bonds connecting sites $(0,0,u)$ and
$(x,y,s)$ as $[x,y,s]\equiv\langle x,y,s|0,0,u\rangle$. Using
(\ref{psg:def}) we can generate any other mean-field bonds through
symmetry operations (such as translations $G_{T_{1,2}}T_{1,2}$ and
mirror reflection $G_\bss\bss$) from $[x,y,s]$. However these
mean-field bonds cannot be chosen arbitrarily since they possess
symmetry relation (\ref{psg:def}):
\begin{eqnarray}
\langle i|j\rangle=G_U(i)~\langle
U^{-1}(i)|U^{-1}(j)\rangle~G^\dagger_U(j)
\end{eqnarray}
where $U$ is any element in the symmetry group. Notice that for time
reversal $\bst$ we have
\begin{eqnarray}
G_\bst(i)\langle i|j\rangle G^\dagger_\bst(j)=-\langle
i|j\rangle\label{psg:def:T}
\end{eqnarray}
We summarize these symmetry conditions on the mean-field bonds
here:

(\rmnum{1}) For $s=u$
\begin{eqnarray}
&\notag \bst:~~~g_\bst[x,y,u]g^\dagger_\bst=-[x,y,u],\\
&\notag T_1^xT_2^{-x}\bss:~~~[x,-x,u]\rightarrow[x,-x,u]^\dagger,\\
&\notag T_1^{x+1}T_2^{y+1}\cs^3:~~~[x,y,u]\rightarrow[x,y,u]^\dagger,\\
&\notag \bss:~~~[x,x,u]\rightarrow[x,x,u].
\end{eqnarray}

(\rmnum{2}) For $s=v$
\begin{eqnarray}
&\notag \bst:~~~g_\bst[x,y,v]g^\dagger_\bst=[x,y,v],\\
&\notag T_2^{y+1}\bss\cs^2:~~~[0,y,v]\rightarrow[0,y,v]^\dagger,\\
&\notag
T_1^{2-2y}T_2^{y-1}\bss\cs^{-1}:~~~[1-2y,y,v]\rightarrow[1-2y,y,v]^\dagger.
\end{eqnarray}

(\rmnum{3}) For $s=w$
\begin{eqnarray}
&\notag \bst:~~~g_\bst[x,y,w]g^\dagger_\bst=[x,y,w],\\
&\notag T_1^{x-1}T_2^{2-2x}\bss\cs:~~~[x,1-2x,w]\rightarrow[x,1-2x,w]^\dagger,\\
&\notag T_1^{x+1}\bss\cs^{-2}:~~~[x,0,w]\rightarrow[x,0,w]^\dagger.
\end{eqnarray}

Now let's consider several simplest examples. At first, on-site
chemical potential terms $\Lambda(x,y,s)=\Lambda_s$ satisfy the
following consistent conditions:
\begin{eqnarray}
&\tau^1\Lambda_s\tau^1=-\Lambda_s;\label{condition:on-site}\\
&\notag g_\bss(u)\Lambda_ug^\dagger_\bss(u)=\Lambda_u,\\
&\notag g_\bss(v)\Lambda_wg^\dagger_\bss(v)=\Lambda_v,\\
&\notag g_\bss(w)\Lambda_vg^\dagger_\bss(w)=\Lambda_w;\\
&\notag g_\cs(u)\Lambda_wg^\dagger_\cs(u)=\Lambda_u,\\
&\notag g_\cs(v)\Lambda_ug^\dagger_\cs(v)=\Lambda_v,\\
&\notag g_\cs(w)\Lambda_vg^\dagger_\cs(w)=\Lambda_w.
\end{eqnarray}
In fact in all 20 $Z_2$ spin on a $\kag$ lattice we all have
$\Lambda_u=\Lambda_v=\Lambda_w\equiv\Lambda_s$ with a proper gauge
choice.

All the 1st n.n. mean-field bonds can be generated from
$u_\alpha\equiv[0,0,v]$. For a generic $Z_2$ spin liquid with PSG
elements $G_\bst(x,y,s)=\imth\tau^1$ and
(\ref{psg:T1,2})(\ref{psg:sig})(\ref{psg:C6}), the bond
$u_\alpha=[0,0,v]$ satisfies the following consistent conditions:
\begin{eqnarray}\label{condition:1st_nn}
&\tau^1u_\alpha\tau^1=-u_\alpha,\\
&g_\bss(u)g_\cs(u)g_\cs(w)u_\alpha
g_\cs^\dagger(v)g_\cs^\dagger(w)g_\bss^\dagger(v)=u_\alpha^\dagger.\notag
\end{eqnarray}
It follows immediately that for six $Z_2$ spin liquids, \ie $\#7-12$
in TABLE \ref{tab:z2kagome} all n.n. mean-field bonds must vanish
since $u_\alpha=0$ as required by (\ref{condition:1st_nn}).
Therefore it's unlikely that the $Z_2$ spin liquid realized in
$\kag$ Hubbard model would be one of these 6 states. In the
following we study the rest 14 $Z_2$ spin liquids on the $\kag$
lattice.

All 2nd n.n. mean-field bonds can be generated from
$u_\beta\equiv[0,1,w]$ which satisfies the following symmetry conditions
\begin{eqnarray}\label{condition:2nd_nn}
&\tau^1u_\beta\tau^1=-u_\beta,\\
&g_\bss(u)g_\cs(u)u_\beta
g_\cs^\dagger(v)g_\bss^\dagger(w)=u_\beta^\dagger.\notag
\end{eqnarray}

There are two kinds of 3rd n.n. mean-field bonds: the first kind can
all be generated by $u_{\gamma}\equiv[1,0,u]$ which satisfies
\begin{eqnarray}\label{condition:3rd_nn1}
&\tau^1u_{\gamma}\tau^1=-u_{\gamma},\\
&g_\cs(u)g_\cs(v)g_\cs(w)u_{\gamma}\big[g_\cs(u)g_\cs(v)g_\cs(w)\big]^\dagger=u_{\gamma}^\dagger.\notag
\end{eqnarray}
the second kind can all be generated by $\tilde
u_{\gamma}\equiv[1,-1,u]$ which satisfies
\begin{eqnarray}\label{condition:3rd_nn2}
&\tau^1\tilde u_{\gamma}\tau^1=-\tilde u_{\gamma},\\
&g_\bss(u)\tilde u_{\gamma}g^\dagger_\bss(u)=\tilde u^\dagger_{\gamma},\notag\\
& g_\cs(u)g_\cs(w)g_\cs(v)\tilde
u_{\gamma}\big[g_\cs(u)g_\cs(w)g_\cs(v)\big]^\dagger=\eta_{12}\tilde
u_{\gamma}^\dagger.\notag
\end{eqnarray}

\section{$Z_2$ spin liquids in the neighborhood of $U(1)$ SL-$[0,\pi]$
state}\label{app:Z2 around U1 Dirac}

\subsection{Mean-field ansatz of $U(1)$ SL-$[0,\pi]$ state}

Following $SU(2)$ Schwinger fermion formulation with
$\psi_i\equiv(f_{i\uparrow},f^\dagger_{i\downarrow})^T$, we focus on
those $Z_2$ spin liquids (SLs) in the neighborhood of $U(1)$
SL-$[0,\pi]$ state with the following mean-field ansatz:
\begin{eqnarray}\label{MF:U1SL0}
&\langle x,y,u|x,y,v\rangle=-\langle
x,y,u|x,y,w\rangle=(-1)^x\chi\tau^3,\\
&\notag\langle x+1,y,w|x,y,u\rangle=\langle
x,y+1,v|x,y,u\rangle=-\langle x,y,v|x,y,w\rangle\\
&\notag=\langle x+1,y-1,w|x,y,v\rangle=\chi\tau^3.
\end{eqnarray}
where $\chi$ is a real hopping parameter. We define mean-field bonds
$\langle x,y,s|x^\prime,y^\prime,s^\prime\rangle$ in the following
way
\begin{eqnarray}
H_{MF}=\sum_{i,j}\psi^\dagger_i\langle i|j\rangle\psi_{j}+~h.c.
\end{eqnarray}
For convenience of later calculation we implement the following
gauge transformation
\begin{eqnarray}
\psi_{x,y,u}\rightarrow\imth\tau^3\psi_{x,y,u}
\end{eqnarray}
and the original mean-field ansatz (\ref{MF:U1SL0}) transforms to be
\begin{eqnarray}\label{MF:U1SL1}
&\langle x,y,u|x,y,v\rangle=-\langle
x,y,u|x,y,w\rangle=\imth(-1)^x\chi\tau^0,\\
&\notag\langle x+1,y,w|x,y,u\rangle=\langle
x,y+1,v|x,y,u\rangle=-\imth\chi\tau^0,\\
&\notag-\langle x,y,v|x,y,w\rangle=\langle
x+1,y-1,w|x,y,v\rangle=\chi\tau^3.
\end{eqnarray}

The projected symmetry group (PSG) corresponds to the above
mean-field ansatz (\ref{MF:U1SL1}) is
\begin{eqnarray}
&\notag G_\bst(x,y,v)=G_\bst(x,y,w)=-G_\bst(x,y,u)=g_\bst,\\
&\notag g_\bst\tau^3 g_\bst^\dagger=-\tau^3;\\
&\notag G_{T_2}(x,y,s)=g_{T_2},~~~g_{T_2}\tau^3g_{T_2}^\dagger
=\tau_3;\\
&\notag
G_{T_1}(x,y,v)=G_{T_1}(x,y,w)=-G_{T_1}(x,y,u)\\
&\notag =(-1)^{x+y}g_{T_1},~~~g_{T_1}\tau^3g_{T_1}^\dagger=\tau_3;\\
&\notag
G_\bss(x,y,v)=G_\bss(x,y,w)=(-1)^{x+y+1}G_\bss(x,y,u)\\
&\notag=(-1)^{(x+y)(x+y+1)/2}g_\bss,~~~g_\bss\tau^3g_\bss^\dagger=\tau^3;\\
&\notag G_{C_6}(x,y,u)=(-1)^{\frac{x(x+1)+y(y-1)}2}g_{C_6},\\
&\notag G_{C_6}(x,y,v)=-(-1)^{\frac{x(x-1)+y(y-1)}2}g_{C_6},\\
&\notag G_{C_6}(x,y,w)=\imth(-1)^{\frac{x(x-1)+y(y-1)}2}g_{C_6}\tau^3,\\
&g_{C_6}\tau^3g_{C_6}^\dagger=\tau^3.\label{psg:U1SL1}
\end{eqnarray}
so that the mean-field ansatz satisfy (\ref{psg:def}).

\subsection{Classification of $Z_2$ spin liquids around $U(1)$
SL-$[0,\pi]$ state}

Plugging (\ref{psg:U1SL1}) into algebraic consistent conditions
(\ref{algebra:psg:T})-(\ref{algebra:psg:C6,T1}) yields four
algebraic solutions of $Z_2$ PSGs around the $U(1)$ SL-$[0,\pi]$
state. Choosing a proper gauge they all satisfy
\begin{eqnarray}\label{psg:all four z2}
&g_\bst=\imth\tau_1,~~~g_{T_1}=g_{T_2}=\tau^0,\\
&\notag \eta_{\bst}=\eta_{12}=\eta_{\cs T_1}=-1,\\
&\notag\eta_{T_{1,2}\bst}=\eta_{\bss T_{1,2}}=\eta_{\cs T_2}=1.
\end{eqnarray}
The four $Z_2$ PSGs near the $U(1)$ SL-$[0,\pi]$ state are featured
by
\begin{eqnarray}\label{psg:z2a}
&(\#6)~Z_2[0,\pi]\alpha:~~~g_\bss=g_\cs=\tau^0,\\
&\notag \eta_\bss=\eta_{\bss\bst}=1,\\
&\notag\eta_{\bss\cs}=\eta_{\cs\bst}=-\eta_\cs=-1;
\end{eqnarray}
\begin{eqnarray}\label{psg:z2b}
&(\#2)~Z_2[0,\pi]\beta:~~~g_\bss=\tau^0,~g_\cs=\imth\tau^3,\\
&\notag \eta_\bss=\eta_{\bss\bst}=1,\\
&\notag\eta_{\bss\cs}=\eta_{\cs\bst}=-\eta_\cs=1;
\end{eqnarray}
\begin{eqnarray}\label{psg:z2c}
&(\#14)~Z_2[0,\pi]\gamma:~~~g_\bss=\imth\tau^3,~g_\cs=\tau^0,\\
&\notag \eta_\bss=\eta_{\bss\bst}=-1,\\
&\notag\eta_{\bss\cs}=\eta_{\cs\bst}=-\eta_\cs=-1;
\end{eqnarray}
\begin{eqnarray}\label{psg:z2d}
&(\#16)~Z_2[0,\pi]\delta:~~~g_\bss=g_\cs=\imth\tau^3\\
&\notag \eta_\bss=\eta_{\bss\bst}=-1,\\
&\notag\eta_{\bss\cs}=\eta_{\cs\bst}=-\eta_\cs=1.
\end{eqnarray}
Of course they belong to the 20 $Z_2$ spin liquids summarized in
TABLE \ref{tab:z2kagome}.

\subsection{Four possible $Z_2$ spin liquids around $U(1)$
SL-$[0,\pi]$ state: mean-field ansatz}

\subsubsection{Consistent conditions on mean-field bonds}

Implementing the generic conditions mentioned earlier on several
near neighbor mean-field bonds with PSG (\ref{psg:all four
z2})-(\ref{psg:z2d}), we obtain the following consistent conditions:

(0) For on-site chemical potential terms
$\Lambda_s(x,y,s)=\vec\lambda(x,y,s)\cdot\vec\tau$, translations
operations $G_{T_{1,2}}T_{1,2}$ in PSG guarantee that
$\Lambda_s(x,y,s)=\lambda_s(0,0,s)\equiv \Lambda_s,~s=u,v,w$. They
satisfy
\begin{eqnarray}
&g_\bst\Lambda_sg^\dagger_\bst=-\Lambda_s;\\
&\notag g_\bss\Lambda_u g_\bss^\dagger=\Lambda_u,~g_\bss\Lambda_v
g_\bss^\dagger=\Lambda_w,~g_\bss\Lambda_w
g_\bss^\dagger=\Lambda_v;\\
&\notag g_\cs\Lambda_u
g_\cs^\dagger=\Lambda_v,~(g_\cs\tau^3)\Lambda_v
(g_\cs\tau^3)^\dagger=\Lambda_w,\\
&\notag g_\cs\Lambda_w g_\cs^\dagger=\Lambda_u.
\end{eqnarray}

(\Rmnum{1}) For 1st neighbor mean-field bond
$u_a\equiv[0,0,v]^\dagger$ (there is only one independent mean-field
bond, meaning all other 1st neighbor bonds can be generated from
[0,0,v] through symmetry operations)
\begin{eqnarray}
&g_\bst u_a^\dagger g_\bst^\dagger=u_a^\dagger,\\
&\notag (g_\bss g_\cs^2\tau^3)u_a^\dagger(g_\bss
g_\cs^2\tau^3)^\dagger=-u_a.
\end{eqnarray}

(\Rmnum{2}) For 2nd neighbor mean-field bond $u_b\equiv[0,1,w]$ we
have
\begin{eqnarray}
&g_\bst u_bg_\bst^\dagger=u_b,\\
&\notag (g_\bss g_\cs)u_b(g_\bss g_\cs)^\dagger=-u_b^\dagger.
\end{eqnarray}

(\Rmnum{2}) For 3rd neighbor mean-field bonds $u_{c1}\equiv[1,0,u]$
and $u_{c2}\equiv[1,-1,u]$ we have
\begin{eqnarray}
&g_\bst u_{c1}g_\bst^\dagger=-u_{c1},\\
&\notag (g_\cs^3\tau^3)u_{c1}(g_\cs^3\tau^3)^\dagger=u_{c1}^\dagger.
\end{eqnarray}
and
\begin{eqnarray}
&g_\bst u_{c2}g_\bst^\dagger=-u_{c2},\\
&\notag g_\bss u_{c2}g^\dagger_\bss=u_{c2}^\dagger,\\
&\notag
(g_\cs^3\tau^3)u_{c2}(g_\cs^3\tau^3)^\dagger=-u_{c2}^\dagger.
\end{eqnarray}

\subsubsection{Mean-field ansatz of the four $Z_2$ spin liquids near
$U(1)$ SL-$[0,\pi]$ state}

For $Z_2[0,\pi]\alpha$ state with $g_\bss=g_\cs=\tau^0$ the
mean-field ansatz are (up to 3rd neighbor mean-field bonds)
\begin{eqnarray}
&u_a=\imth a_0\tau^0+a_1\tau^1,~~~u_b=\imth b_0\tau^0,\\
&\notag u_{c1}=c_3\tau^3,~~~u_{c2}=c_2\tau^2,\\
&\notag \Lambda_s=\lambda_3\tau^3,~~~s=u,v,w.
\end{eqnarray}
Since we are considering a phase perturbed from the $U(1)$
SL-$[0,\pi]$ state, we shall always assume $a_0\neq0$ (1st neighbor
hopping terms) in the following discussion. A $Z_2[0,\pi]\alpha$
spin liquid can be realized by 1st neighbor mean-field singlet
pairing terms with $a_1\neq0$.

For $Z_2[0,\pi]\beta$ state with $g_\bss=\tau^0,~g_\cs=\imth\tau^3$
the mean-field ansatz are (up to 3rd neighbor mean-field bonds)
\begin{eqnarray}
&u_a=\imth a_0\tau^0+a_1\tau^1,~~~u_b=\imth b_0\tau^0+b_1\tau^1,\\
&\notag u_{c1}=c_2\tau^2+c_3\tau^3,~~~u_{c2}=0,\\
&\notag
\Lambda_u=\lambda_2\tau^2+\lambda_3\tau^3,~\Lambda_{v,w}=-\lambda_2\tau^2+\lambda_3\tau^3.
\end{eqnarray}
A $Z_2[0,\pi]\beta$ spin liquid can be realized by 2nd neighbor
pairing terms with $a_0b_1-a_1b_0\neq0$.

For $Z_2[0,\pi]\gamma$ state with $g_\bss=\imth\tau^3,~g_\cs=\tau^0$
the mean-field ansatz are (up to 3rd neighbor mean-field bonds)
\begin{eqnarray}
&u_a=\imth a_0\tau^0,~~~u_b=\imth b_0\tau^0+b_1\tau^1,\\
&\notag u_{c1}=c_3\tau^3,~~~u_{c2}=0,\\
&\notag \Lambda_s=\lambda_3\tau^3,~~~s=u,v,w.
\end{eqnarray}
A $Z_2[0,\pi]\gamma$ spin liquid can be realized by 2nd neighbor
pairing terms with $b_1\neq0$.

For $Z_2[0,\pi]\delta$ state with $g_\bss=g_\cs=\imth\tau^3$ the
mean-field ansatz are (up to 3rd neighbor mean-field bonds)
\begin{eqnarray}
&u_a=\imth a_0\tau^0,~~~u_b=\imth b_0\tau^0,\\
&\notag u_{c1}=c_2\tau^2+c_3\tau^3,~~~u_{c2}=0,\\
&\notag \Lambda_s=\lambda_3\tau^3,~~~s=u,v,w.
\end{eqnarray}
A $Z_2[0,\pi]\delta$ spin liquid can be realized by 3rd neighbor
pairing terms with $c_2\neq0$.

\subsection{Low-energy effective theory}

The reciprocal unit vectors (corresponding to unit vectors $\vec
a_{1,2}$) on a $\kag$ lattice are $\vec b_1=\frac1a(\hat
x-\frac1{\sqrt3}\hat y)$ and $\vec b_2=\frac1a\frac2{\sqrt3}\hat y$,
satisfying $\vec a_i\cdot\vec b_j=\delta_{i,j}$. In the mean-field
ansatz (\ref{MF:U1SL1}) of $U(1)$ SL-$[0,\pi]$ the unit cell is
doubled whose translation unit vectors are $\vec A_1=2\vec a_1$ and
$\vec A_2=\vec a_2$. Accordingly the 1st BZ for such a mean-field
ansatz is only half of the original 1st BZ with new reciprocal unit
vectors being $\vec B_1=\vec b_1/2$ and $\vec B_2=\vec b_2$.
Denoting the momentum as $\tk\equiv (k_x,k_y)/a=k_1\vec B_1+k_2\vec
B_2$ with $|k_{1,2}|\leq\pi$, we have
\begin{eqnarray}
k_1=2k_x,~~~k_2=(k_x+\sqrt3k_y)/2.
\end{eqnarray}
The two Dirac cones in the spectra of $U(1)$ SL-$[0,\pi]$ state
(\ref{MF:U1SL1}) are located at $\pm\textbf{Q}$ with
\begin{eqnarray}
\textbf{Q}=(0,\frac{\pi}{\sqrt3})=\frac{\pi}2\vec B_2
\end{eqnarray}
with the proper chemical potential $\Lambda(i)=\langle
i|i\rangle=\chi(\sqrt3-1)\tau^3$ added to mean-field ansatz
(\ref{MF:U1SL1}).

For convenience we choose the following basis for Dirac-like
Hamiltonian obtained from expansion around $\pm\textbf{Q}$:
\begin{eqnarray}
&\notag\phi_{+,\uparrow,A}=\frac1{\sqrt6}e^{-\imth\frac1{24}\pi}\cdot\\
&\notag(e^{-\imth\frac{11}{12}\pi},0,e^{\imth\frac{11}{12}\pi},0,0,0,e^{-\imth\frac{11}{12}\pi},0,e^{\imth\frac{5}{12}\pi},0,\sqrt2,0)^T,\\
&\notag\phi_{+,\uparrow,B}=\frac1{\sqrt6}e^{-\imth\frac1{24}\pi}\cdot\\
&\notag(1,0,e^{-\imth\frac43\pi},0,\sqrt2e^{-\imth\frac{11}{12}\pi},0,-1,0,e^{-\imth\frac{5}{6}\pi},0,0,0)^T,\\
&\notag\phi_{-,\uparrow,b}=R_{T_1}(k_1=0,k_2=-\frac\pi2)\phi_{+,\uparrow,b},\\
&\label{dirac:basis}\phi_{\pm,\downarrow,b}=R_T\phi_{\pm,\uparrow,b}.
\end{eqnarray}
where $\pm$ are valley index for two Dirac cones at $\pm\textbf{Q}$
with Pauli matrices $\boldsymbol{\mu}$ and $b=A,B$ are band indices (for the two bands forming the Dirac cone)
with Pauli matrices $\boldsymbol{\nu}$. Pseudospin indices
$\Sigma=\uparrow,\downarrow$ are assigned to the two degenerate bands related by time reversal, with Pauli matrices
$\boldsymbol{\sigma}$. The corresponding creation operators for
these modes are
$\Psi^\dagger_{\pm,\Sigma,b}=\psi_{\pm\textbf{Q}}^\dagger\phi_{\pm,\Sigma,b}$
in the order of $(0,0,u),(0,0,v),(0,0,w),(1,0,u),(1,0,v),(1,0,w)$
for the six sites per doubled new unit cell. Notice that in terms of $f$-spinons we have $\psi^\dagger=(f^\dagger_\uparrow,f_\downarrow)$.

Here $R_\bst\equiv
I_{2\times2}\otimes\begin{bmatrix}-1&0&0\\0&1&0\\0&0&1\end{bmatrix}\otimes
g_\bst$,~$R_{T_2}(\tk)=e^{-\imth k_2}I_{6\times6}\otimes g_{T_2}$
and $R_{T_1}(\tk)=\begin{bmatrix}0&-e^{-\imth
k_1}\\1&0\end{bmatrix}\otimes\begin{bmatrix}1&0&0\\0&-1&0\\0&0&-1\end{bmatrix}\otimes
g_{T_1}$ are transformation matrices on 12-component eigenvectors
for time reversal $\bst$ and translation $T_{1,2}$ operations. By
definition of PSG the eigenvectors $\phi_\tk$ with momentum
$\tk=k_1\vec B_1+k_2\vec B_2\equiv(k_1,k_2)$ and energy $E$ have the
following symmetric properties:
\begin{eqnarray}
&\notag\bst:~~~\tilde\phi_{(k_1,k_2)}(E)=R_\bst\phi_{(k_1,k_2)}(-E),\\
&\notag T_1:~~~\tilde\phi_{(k_1,k_2)}(E)=R_{T_1}(k_1,k_2)\phi_{(k_1,k_2+\pi)}(E),\\
&\notag
T_2:~~~\tilde\phi_{(k_1,k_2)}(E)=R_{T_2}(k_1,k_2)\phi_{(k_1,k_2)}(E).
\end{eqnarray}
$\tilde\phi$ and $\phi$ are the basis after and before the symmetry
operations.

In such a set of basis the Dirac Hamiltonian obtained by expanding
the $U(1)$ SL-$[0,\pi]$ mean-field ansatz (\ref{MF:U1SL1}) around
the two cones at $\pm\textbf{Q}$ is
\begin{eqnarray}
H_{\text{Dirac}}=\sum_\tk\frac\chi{\sqrt2}\Psi_\tk^\dagger\mu^0\sigma^3(-k_x\nu^1+k_y\nu^2)\Psi_\tk
\end{eqnarray}
$\tk$ should be understood as small momenta measured from
$\pm\textbf{Q}$. Possible mass terms are
$\mu^{0,1,2,3}\sigma^{1,2}\nu^0$ and
$\mu^{0,1,2,3}\sigma^{0,3}\nu^3$. However not all of them are
allowed by symmetry. Here we numerate all symmetry operations and
associated operator transformations:

Spin rotation along $\hat z$-axis by angle $\theta$:
\begin{eqnarray}
\notag\Psi^\dagger_\tk\rightarrow\Psi^\dagger_\tk
e^{\imth\frac\theta2}
\end{eqnarray}
Spin rotation along $\hat y$-axis by $\pi$:
\begin{eqnarray}
\notag\Psi_\tk^\dagger\rightarrow\Psi^T_{-\tk}\mu^2\sigma^2\nu^2
\end{eqnarray}
Time reversal $\bst$:
\begin{eqnarray}
\notag\Psi_\tk^\dagger\rightarrow\Psi^\dagger_\tk(-\imth\sigma^2)
\end{eqnarray}
Translation $T_1$:
\begin{eqnarray}
\notag\Psi_\tk^\dagger\rightarrow\Psi^\dagger_\tk(-\mu^3)
\end{eqnarray}
Translation $T_2$:
\begin{eqnarray}
\notag\Psi_\tk^\dagger\rightarrow\Psi^\dagger_\tk(-\imth\mu^3)
\end{eqnarray}
Considering the above conditions, the only symmetry-allowed mass
terms are $\sum_\tk\Psi^\dagger_\tk m_{1,2}\Psi_\tk$ with
$m_1=\mu^0\sigma^1\nu^0$ and $m_2=\mu^3\sigma^3\nu^3$.

The transformation rules for mirror reflection $\bss$ and $\pi/3$
rotation $C_6$ depend on the choice of $g_\bss,g_\cs$ in the PSG. In
general we have
\begin{eqnarray}
&\notag\bss:~~~\Psi^\dagger_{\tk}\rightarrow\Psi^\dagger_{\bss\tk}M_\bss(g_\bss),\\
&\notag\cs:~~~\Psi^\dagger_{\tk}\rightarrow\Psi^\dagger_{\cs\tk}M_\cs(g_\cs).
\end{eqnarray}
Using the basis (\ref{dirac:basis}) the $8\times8$ matrices
$M_{\bss,\cs}$ can be expressed in terms of Pauli matrices
$\boldsymbol{\mu}\otimes\boldsymbol{\sigma}\otimes\boldsymbol{\nu}$.
For the four $Z_2$ spin liquid we have
\begin{eqnarray}
&\notag M_\bss(g_\bss=\tau^0)=\mu^3\otimes\sigma^0\otimes\begin{pmatrix}0&e^{-\imth\frac1{12}\pi}\\e^{-\imth\frac5{12}\pi}&0\end{pmatrix},\\
&\notag M_\bss(g_\bss=\imth\tau^3)=\mu^3\otimes\sigma^3\otimes\begin{pmatrix}0&e^{\imth\frac5{12}\pi}\\e^{\imth\frac1{12}\pi}&0\end{pmatrix};\\
&\notag
M_\cs(g_\cs=\tau^0)=\begin{pmatrix}1&0\\0&\imth\end{pmatrix}\otimes\sigma^0\otimes
e^{\imth\frac76\pi\nu^3},\\
&\notag
M_\cs(g_\cs=\imth\tau^3)=\begin{pmatrix}\imth&0\\0&-1\end{pmatrix}\otimes\sigma^0\otimes
e^{\imth\frac16\pi\nu^3}.
\end{eqnarray}

It turns out in $Z_2[0,\pi]\beta$ state, only the 1st mass term
$m_1=\mu^0\sigma^1\nu^0$ is invariant under $\bss$ and $C_6$
operations. In other 3 states neither mass terms $m_{1,2}$ are
symmetry-allowed. As a result we only have one gapped $Z_2$ spin
liquid, \ie $Z_2[0,\pi]\beta$ state in the neighborhood of $U(1)$
Dirac SL-$[0,\pi]$ state.

Let's consider mean-field bonds up to 2nd neighbor for ansatz
$Z_2[0,\pi]\beta$. Perturbations to the two Dirac cones of $U(1)$
SL-$[0,\pi]$ with $\lambda_3=(\sqrt3-1)a_0$ in general has the
following form
\begin{eqnarray}
&\notag\delta H_0=\big[\lambda_3-(\sqrt3-1)
a_0-(\sqrt3+1)b_0\big]\mu^0\sigma^3\nu^0\\
&+\big[(\sqrt3+1)b_1-\lambda_2-(\sqrt3-1)a_1\big]\mu^0\sigma^1\nu^0
\end{eqnarray}
This means we need either 1st neighbor ($a_1$) or 2nd neighbor
($b_1$) pairing term to open up a gap in the spectrum. Meanwhile
these pairing terms break the original $U(1)$ symmetry down to $Z_2$ symmetry.


\section{$Z_2$ spin liquids in the neighborhood of uniform RVB
state}\label{app:Z2 around u-RVB}

The mean-field ansatz of the uniform RVB state is simple:
\begin{eqnarray}
H_{MF}=\chi\sum_{<ij>,\sigma}f^\dagger_{i,\sigma}f_{j,\sigma}
\end{eqnarray}
where $\chi$ is a real parameter and $<ij>$ represents sites $i,j$
being nearest neighbor (n.n.) of each other. It's straightforward to
show the PSG elements of such a mean-field ansatz are
\begin{eqnarray}\label{psg:uRVB}
G_U(x,y,s)=g_U,~~~U=T_{1,2},~\bst,~\bss,~\cs.
\end{eqnarray}
and $SU(2)$ matrices $g_U$ satisfy
\begin{eqnarray}
&g_\bst\tau^3g_\bst^\dagger=-\tau^3,\\
&\notag g_U\tau^3g_U^\dagger=\tau^3,~~~U=T_{1,2},\bss,\cs.
\end{eqnarray}
It turns out there are only 4 gauge-inequivalent $Z_2$ PSGs as
solutions to (\ref{algebra:psg:T})-(\ref{algebra:psg:C6,T2}) with
the form (\ref{psg:uRVB}). In other words, there are only 4
different $Z_2$ in the neighborhood of a uniform RVB states.
Choosing a proper gauge they all satisfy $g_\bst=\imth\tau^1$,
$g_{T_{1,2}}=\tau^0$ and $\eta_{T_{1,2}\bst}=\eta_{12}=\eta_{\cs
T_{1,2}}=\eta_{\bss T_{1,2}}=1,~\eta_\bst=-1$. These four states are
characterized by:
\begin{eqnarray}
&\label{z2A}(\#1)~Z_2[0,0]A:~~~g_\bss=g_\cs=\tau^0,\\
&\notag\eta_{\bss\bst}=\eta_{\cs\bst}=\eta_{\bss}=\eta_\cs=\eta_{\bss\cs}=1.
\end{eqnarray}
\begin{eqnarray}
&\label{z2B}(\#5)~Z_2[0,0]B:~~~g_\bss=\tau^0,~g_\cs=\imth\tau^3,\\
&\notag\eta_{\bss\bst}=\eta_{\bss}=1,~\eta_{\cs\bst}=\eta_\cs=\eta_{\bss\cs}=-1.
\end{eqnarray}
\begin{eqnarray}
&\label{z2C}(\#15)~Z_2[0,0]C:~~~g_\bss=\imth\tau^3,~g_\cs=\tau^0,\\
&\notag\eta_{\bss\bst}=\eta_{\bss}=-1,~\eta_{\cs\bst}=\eta_\cs=\eta_{\bss\cs}=1.
\end{eqnarray}
\begin{eqnarray}
&\label{z2D}(\#13)~Z_2[0,0]D:~~~g_\bss=g_\cs=\imth\tau^3,\\
&\notag\eta_{\bss\bst}=\eta_{\cs\bst}=\eta_{\bss}=\eta_\cs=\eta_{\bss\cs}=-1.
\end{eqnarray}
It turns out these four $Z_2$ SLs around uniform RVB state are all gapped as shown in TABLE \ref{tab:z2kagome}.\\


\end{document}